\documentclass[a4paper]{article}

\usepackage{fullpage}
\usepackage[braket, qm]{qcircuit}
\usepackage{braket}
\usepackage{amsmath}
\pdfmapfile{+sansmathaccent.map}
\usepackage{dsfont}
\usepackage{graphicx}
\usepackage{hyperref}
\usepackage{tikz}
\usepackage{caption}
\usepackage{authblk}
\usetikzlibrary{positioning,shapes,shadows,arrows}

\date{March 9, 2021}
\title{A Quantum Algorithm for the Sensitivity Analysis\\ of Business Risks}
\author[1]{M.C. Braun$^*$}
\author[1]{T. Decker\footnote{For contact email: thomas.decker@jos-quantum.de or sven.kerstan@jos-quantum.de. Authors are listed in alphabetical order.}}
\author[1]{N. Hegemann$^*$}
\author[1]{S.F. Kerstan$^*$}
\author[2]{C. Schäfer$^*$}
\affil[1]{JoS QUANTUM GmbH\\Frankfurt am Main\\Germany}
\affil[2]{Deutsche Börse Group\\Frankfurt am Main\\Germany}

\begin{document}

\maketitle

\abstract{We present a novel use case for quantum computation: the sensitivity analysis for a risk model used at Deutsche Börse Group. Such an analysis is computationally too expensive to perform on classical computers. We show in detail how the risk model and its analysis can be implemented as a quantum circuit. We test small scale versions of the model in simulation and find that the expected quadratic speedup compared to the classical implementation used at Deutsche Börse Group can be realized. Full scale production usage would be possible with less than 200 error corrected qubits.\\
Our quantum algorithm introduces unitary but imperfect oracles which use Quantum Amplitude Estimation to detect and mark states. This construction should be of general interest and we present theoretical results regarding the performance of Grover's search algorithm with imperfect oracles.

\section{Introduction}
Quantum computers promise to speed up some calculations which would require an impractical amount of calculation time on classical computers. Readers looking for a technical introduction to the subject will find reference \cite{nielsenchuang} useful, although it does not cover the most recent developments that matter for this work, in particular reference \cite{QAE}.\\
The best known quantum algorithms either offer a super-polynomial speedup over known classical implementations like Shor's algorithm \cite{shor} or a quadratic speedup like Grover's search \cite{grover,grover2,bbht98}, which will play an important role for our application. Many other quantum algorithms are variations or extensions of the ideas of these algorithms and they exhibit the same speedups. An important example of an extension of Grover's algorithm is Quantum Amplitude Estimation ("QAE") \cite{QAE}, which also plays an important role in the quantum implementation of our use case.\\
Despite the progress regarding algorithms and hardware, very few concrete business problems, which can be solved with relevant speed gains on quantum computers, have been described in the literature, especially for applications in finance. An example is \cite{quantumriskanalysis}, where a quantum speedup for Value at Risk and related calculations is demonstrated. Note that classical Quasi-Monte Carlo methods will often scale (almost) as well as the quantum solution, as long as the problem is not high dimensional \cite{fries1},\cite{fries2}. Another application described in the literature regards option pricing, \cite{quantumpricing0}, \cite{quantumoptionpricing}. The methods described there should turn out to be useful for some exotic options, even though certain examples mentioned there (Barrier options and Asian options) are in some cases already priced very efficiently with standard, non-stochastic methods by practitioners in the industry. For other path dependent options, which correspond to high dimensional problems, quantum computation should outperform classical quasi-Monte Carlo methods.\\
In this note, we present another example -- the sensitivity analysis of a business risk model at Deutsche Börse Group -- and we implement and test the performance of it for a small scale model in simulation. We consider this as interesting for three reasons: First, a classical implementation cannot deliver results fast enough to be of practical use, while the quantum version should be able to do so as soon as the necessary quantum hardware is available. So the quantum implementation would have practical value for a real-world problem. Second, the requirements regarding the number of qubits seems to be low -- in the order of 200 qubits for the full model -- and may be met by commercially available quantum systems within a few years of time. Third, our quantum program has an interesting structure: The quantum risk model sits inside the QAE, and the QAE sits inside the Grover algorithm. We expect that quantum programs with structures like this are what will enter productive use within a few years time, rather than setups where just a simple Grover search or QAE is used in an otherwise classical environment.\\
We mentioned that in our quantum algorithm, QAE sits inside the Grover algorithm. To be more precise: The QAE is used to calculate states that correspond to binary encodings of the desired result. The most important states that are closest to the solution and lead to a success probability of at least $8/\pi^2$ of the QAE (see theorem 12 in \cite{bbht98}) are used to control a phase operation. This construction leads to an imperfect Grover oracle and such a construction raises some questions regarding the theory, since our oracles do not mark the relevant states with certainty. Given that kind of oracle, does the Grover search still work as intended? And if so, what is the number of Grover steps that yield the maximal success probability for the algorithm? And what is the maximal success probability? We address these questions in section \ref{theory}.\\
Grover search with imperfect, non-unitary oracles has been analyzed in \cite{faultygrover}. In this work, it was shown that the quantum speedup is destroyed by a faulty, non-unitary oracle. Fortunately this does not apply to our case, since our oracles are unitary. We show that the oracles we consider lower the success probability by a constant factor compared to standard oracles. In our application, we achieve a success probability of at least 81\%, rather than close to 100\% as in conventional Grover, which is inherited from the QAE. The number of optimal search steps remains similar compared to the original Grover algorithm and causes no problem in practice.\\
Imperfect, unitary oracles have been discussed in \cite{walksearch} as approximate reflection operators in the context of search algorithms based on quantum walks (see also for example \cite{akr}). Even though these seem to have some similarity with our theoretical results, there and also in the more general literature on variations of Grover's original quantum search algorithm (e.g., see \cite{ambianis} and \cite{tulsi}) it appears that our result has not been described before.\\
The structure of the paper is as follows: In section~\ref{usecase}, the classical business risk model used at Deutsche Börse Group is described, with a concrete, small scale example. In section~\ref{quantum}, we show how the example and its sensitivity analysis is implemented as a quantum program and we run it in simulation. Here, as in the rest of the paper, we use Qiskit~\cite{qiskit} to run the code and present Qiskit style circuit diagrams. Section~\ref{scalingexperiments} shows how our quantum program scales with an increasing (but very small) size of the model in simulations. In section~\ref{outlook} we sketch the requirements that quantum hardware would have to meet to run our quantum program full scale. In section \ref{theory} we come back to the questions about the fractional oracles we used in our algorithm. We give formulas for the optimal number of steps and for the maximum success probability for Grover with our oracles (in practice, the maximum is what is (almost) achieved, the minimum is typically only relevant for toy models). In appendix~\ref{proof}, we present the proofs for the formulas given in section \ref{theory}.

\section{Use case: Business risk analysis} \label{usecase}
An important question for any business is: What would be the impact of external adverse developments on future revenues? For example, these developments could be caused by 
(macro-)economic, or political events, by competition, or changes in the company's reputation, or regulation or tax, etc. A major problem for the analysis of this kind of environmental circumstances is that these factors not only (potentially) influence revenues, but are also interlinked in many cases, i.e. they influence each other.\\
One can build a risk model to quantify such scenarios and their impact on revenues. This allows for estimating the overall likelihood of impacts that would threaten the business. If this likelihood is too high for the risk appetite of the company, then we should take action, for example by trying to hedge risks, diversifying or pulling out of certain markets.\\
To make this more concrete, we define a threshold $A$ for the financial impact. Then we define the probability $P(A)$ that this threshold is breached, i.e., the probability that we will see a financial impact bigger than $A$. Next, we define a maximal acceptable value for $P(A)$, which we call $P_{\rm max}$. Once we have a business risk model and estimated the inputs, we can calculate $P(A)$ and check if all is well, that is, if $P(A) < P_{\rm max}$. Should the risk model show that the probability of a breach is bigger than $P_{\rm max}$, then this indicates that action needs to be taken.\\
So let us suppose all the inputs for the risk models have been estimated and the model does show that $P(A)<P_{\rm max}$, so all seems well. But since there is uncertainty in all of the estimated inputs, we would like to know: Is there a parameter (or a combination of parameters) which, when changed slightly, changes the results of the model such that $P(A) > P_{\rm max}$? In other words: Is our result robust? Currently, the large computational effort required to answer this question makes it impossible to analyze this issue thoroughly.

\subsection{The business risk model at Deutsche Börse Group}
A quantitative treatment of business risks has been implemented at Deutsche Börse Group as follows:
Each relevant event is called a risk item (RI, where ${\rm RI}_i$ denotes the $i$-th risk item) and it is assigned an intrinsic probability $p_i$ of occurring without an explicit root cause. An example could be a major down or upward movement in the stock market. On top of that, the $i$-th RI can also be assigned a probability to trigger one or more related RIs, say the $j$-th RI, with a transition probability $p_{ij}$. A major move in the stock market might, for example, trigger a significant change in trading volumes on the exchanges.
Therefore, the model has the structure of a tree, as in the small scale example in figure~\ref{fig:figure1}:
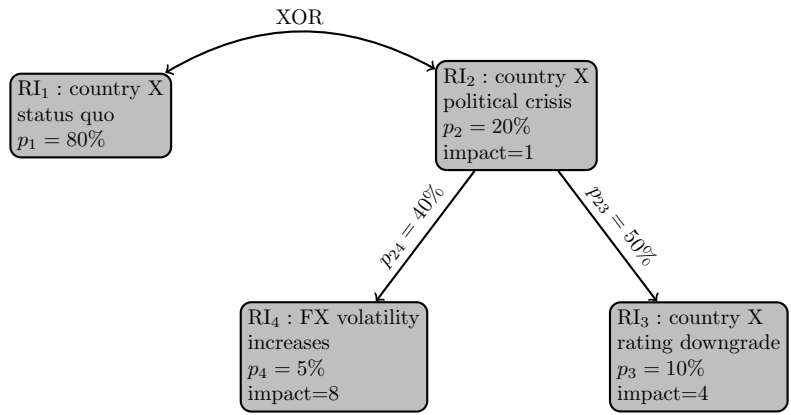
\begin{figure}[h]
\centering
\captionsetup{font=small, width=0.8\textwidth}
\begin{tikzpicture}[thick,scale=0.8, every node/.style={transform shape}]
\node[rectangle,rounded corners,draw=black,align=left,fill=lightgray] (statusquo) at (2,0) {${\rm RI}_1:$ country X\\ status quo\\$p_1=80\%$};
\node[rectangle,rounded corners,draw=black,align=left,fill=lightgray] (origin) at (9, 0) {${\rm RI}_2:$ country X\\ political crisis\\ $p_2=20\%$\\impact=1};
\node[rectangle,rounded corners,draw=black,align=left,fill=lightgray] (sales) at (12, -4) {${\rm RI}_3:$ country X\\ rating downgrade\\$p_3=10\%$\\impact=4};
\node[rectangle,rounded corners,draw=black,align=left,fill=lightgray] (crash) at (6, -4) {${\rm RI}_4:$ FX volatility\\ increases \\$p_4=5\%$\\impact=8};
\path[<->,draw,thick] (statusquo) edge[bend left] node[above] {XOR} (origin);
\path[->,draw,thick] (origin) edge node[sloped, above] {$p_{24}=40\%$} (crash);
\path[->,draw,thick] (origin) edge node[sloped, above] {$p_{23}=50\%$} (sales);
\end{tikzpicture}
\captionof{figure}{Modeling potential losses in a business risk model as used by Deutsche Börse Group. In this example, a country X may experience a political crisis (${\rm RI}_2$). If this happens, it has a 50\% chance of triggering a rating downgrade for country X (${\rm RI}_3$). The downgrade may also happen without a political crisis, with a probability of 10\%. A crisis in country X may also trigger an increased exchange rate (FX) volatility (${\rm RI}_4$), with a probability of 40\%. Again, the FX volatility can also increase without a political crisis with a probability of 5\%. Each item comes with an impact, which should be thought of as a loss of revenues.}\label{fig:figure1}
\end{figure}

In a final step, each RI, which was triggered either intrinsically or by a transition from another triggered RI, generates a specific loss.
The loss for a specific scenario is the sum of the losses of the triggered RIs. \\
The model can then be evaluated by brute force: One draws a random number uniformly between $0$ and $1$ for each risk item and for each transition probability. If the number is smaller than the probability determined in the model, then the RI or transition is triggered. This way, a cascading effect can trigger chain reactions.\\
To illustrate how the evaluation of the model is done, we draw five random numbers (the exclusive-or structure between ${\rm RI}_1$ and ${\rm RI}_2$ means we need to draw only for one of these, the other is then fixed) and determine the loss associated to the scenario represented by these random numbers in the table below.

\begin{table}[h!]
  \begin{center}
    \begin{tabular}{|l|c|c|c|c|c|}
      \textbf{} & \textbf{probability} & \textbf{random number} & \textbf{intr. trig.} & \textbf{trig. by trans.} & \textbf{loss}\\
      \hline
      ${\rm RI}_1$ & $0.80$ & $0.91$ 		& no	& no & $0$\\
      ${\rm RI}_2$ & $0.20$ & $1.00-0.91=0.09$ 	& yes	& no& $1$\\
      ${\rm RI}_3$ & $0.10$ & $0.41$ 		& no	& no& $0$\\
	  ${\rm RI}_4$ & $0.05$ & $0.87$ 		& no	& yes& $8$\\
      ${\rm T}_{23}$  & $0.50$ & $0.51$			& no	& --& --\\
      ${\rm T}_{24}$  & $0.40$ & $0.11$ 		& yes 	& --& --\\
    \end{tabular}
  \end{center}
\end{table}

The exclusive-or structure between ${\rm RI}_1$ and ${\rm RI}_2$ means that regardless of the random numbers exactly one of ${\rm RI}_1$ and ${\rm RI}_2$ is triggered.
With the random numbers we drew, ${\rm RI}_2$ and the transition from ${\rm RI}_2$ to ${\rm RI}_4$ are triggered, but nothing else. This means that ${\rm RI}_4$ is triggered regardless of the random number drawn for the intrinsic trigger of ${\rm RI}_4$. The loss generated by this scenario is $9$ units.\\
When the model is evaluated this way many times, we can think of the process as a Monte Carlo simulation of the model, and this allows us to create a probability distribution for the losses.\\
This loss distribution is then evaluated to determine a certain percentile of the losses, e.g. 99.5\%, so that only the remaining 0.5\% of the worst case losses exceed the respective loss amount; in other words: one can be sure at a 99.5\% confidence level that the loss will not be larger than this percentile of the loss distribution. Note that in our quantum implementation in the following section, we will turn this around and set the maximal acceptable financial impact, rather than the percentage, as explained earlier in this section. We will choose the maximal acceptable impact $A$ to be 12 units and calculate the probability $P(A)$ of breaching $A$.\\
Since the slowness and probabilistic results are disadvantages to Monte Carlo methods, an obvious question is if the model can be solved analytically. The structure of the problem makes that very difficult. To illustrate that, we think of the model as a tree. Then the fact that each risk item can either be triggered intrinsically or by another introduces a path dependency in the model for connected risk items. That means that the number of terms needed to describe the model analytically should grow exponentially in the number of risk items. The XOR structure further complicates the matter. An analytical solution seems to be out of reach.\\
In practice, a brute force Monte Carlo simulation is slow but it is still practical to evaluate the full risk model with several hundred parameters. However, since many of the model parameters, i.e.,the probabilities for RIs and transitions, are estimates, it would be very useful to understand how sensitive the results are with respect to changes in the estimates. In particular, it might be the case that changing the value of one parameter by a few percent might lead to significantly worse outcomes, i.e., the share of events exceeding the previously calculated percentile might exceed the acceptable 0.5\% quota.\\
For robust risk management, it would therefore be useful to be able to identify parameters, which have a very large effect on the results. One could then either try to improve the estimates for just those parameters (the effort required to do this for all parameters might be too big), or simulate worst-case scenarios for a range of values for that parameter.\\
Due to the sheer size of the model currently implemented at Deutsche Börse Group with roughly 400 parameters, it would require 400 times the effort of running the model to test the sensitivity to each input. With the current setup, this would require days or even weeks of calculation time. Analyzing the impact of varying all possible pairs of parameters would require almost 80,000 times the effort and require years of time, and triplets in the order ten million times the effort, requiring centuries of calculation time.

\section{Quantum implementation}\label{quantum}
Now we implement the sensitivity analysis of the risk model as a quantum program. The program analyzes the impact of varying each input parameter separately. The quantum algorithm consists of three steps:
\begin{enumerate}
\item{Implement the risk model as a quantum algorithm.}
\item{Implement QAE on the outputs of the risk model.}
\item{Search sensitive parameters with Grover's algorithm.}
\end{enumerate}
Note that the result from the second step is interesting in its own right, since it corresponds to the quantum version of what is currently done classically at Deutsche Börse Group.\\
In this section, we will go through each of these three steps in detail with the specific example from section \ref{usecase} including a number of circuit diagrams.

\subsection{Quantum circuit for the business risk model}\label{themodel}
The first step is relatively straightforward since the structure of the model translates directly into a quantum circuit. For the concrete example from section \ref{usecase}, which had just four risk items, a quantum program built from elementary gates would look like the circuit in figure~\ref{baremodel}.
\begin{figure}[h]
\centering
\captionsetup{font=small,width=0.8\textwidth}
\scalebox{0.8}
{
\Qcircuit@C=0.5em@R=0.5em@!R
{\lstick{{q}_{0}:}&\qw&\qw&\qw&\qw&\qw&\qw&\qw&\gate{U3(0.451)}&\qw&\gate{U3(1.430)}&\qw&\qw\\
\lstick{{q}_{1}:}&\qw&\qw&\qw&\gate{U3(0.643)}&\qw&\gate{U3(1.671)}&\qw&\qw&\qw&\qw&\qw&\qw\\
\lstick{{q}_{2}:}&\qw&\qw&\targ&\ctrlo{-1}&\qw&\ctrl{-1}&\qw&\ctrlo{-2}&\qw&\ctrl{-2}&\qw&\qw\\
\lstick{{q}_{3}:}&\gate{U3(2.214)}&\qw&\ctrlo{-1}&\qw&\qw&\qw&\qw&\qw&\qw&\qw&\qw&\qw}
}
\caption{The quantum circuit diagram for the simplified risk model described in section \ref{usecase} with 4 risk items. With the notation from Qiskit, the $U3$ gates have three parameters, but we left out the the two phase parameters $\phi$ and $\lambda$ because both are $0$ in our circuit.}
\label{baremodel}
\end{figure}
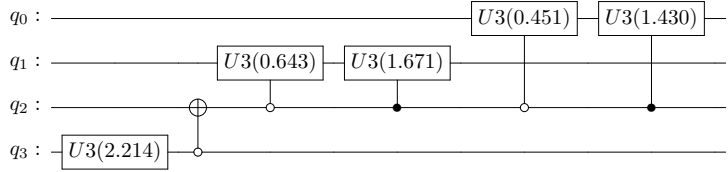
The circuit is initialized with all four qubits in state $|0\rangle$. The state $|0\rangle$ represents a RI that is not triggered and the state $|1\rangle$ a RI that is triggered. Each of the four qubits in the circuit represents one of the four risk items in the example from section \ref{usecase}. The risk items ${\rm RI}_1$, ${\rm RI}_2$, ${\rm RI}_3$, ${\rm RI}_4$ are represented by qubits $q_3, q_2, q_1, q_0$, respectively. We can now put a risk item into a superposition of being triggered with probability $P$ and not being triggered with probability $1-P$ simply by performing a rotation 
$$
U3(\theta,\phi,\lambda)=\left(\begin{array}{cc}{\rm cos}(\theta/2)&-e^{i\lambda}{\rm sin}(\theta/2)\\e^{i\phi}{\rm sin}(\theta/2)&e^{i(\phi+\lambda)}{\rm cos}(\theta/2)\end{array}\right)
$$
on a qubit. The angle $\theta$ must be chosen such that 
\begin{equation}
{\rm sin}^2(\theta/2)=P \label{angle}.
\end{equation}
There is an additional degree of freedom, since we could use non-trivial phases $\phi$ and $\lambda$ in each $U3(\theta, \phi, \lambda)$ gate and still measure the 
same probabilities for each risk items. We choose the phases to be $0$.
For example, if we rotate qubit $q_3$, which represents ${\rm RI}_1$, by $\theta=2.214$ radians away from its initial state $\ket{0}$, then we will measure $q_3$ to be in state $\ket{1}$ at the end of the circuit with probability $P=80\%$, and in state $\ket{0}$ with probability $1-P=20\%$. So this is the quantum implementation of ${\rm RI}_1$ which is to be triggered with eighty percent likelihood in the classical model.\\
Next, we look at the exclusive-or structure between ${\rm RI}_1$ and ${\rm RI}_2$ in the classical model. This can simply be implemented by a controlled NOT gate between $q_2$ and $q_3$.\\
Finally, since ${\rm RI}_3$ and ${\rm RI}_4$ can be triggered by ${\rm RI}_2$ via the transition probability, the corresponding qubits $q_1$ and $q_0$ are each set up with two controlled rotations: The first rotation is executed if the controlling qubit is in state $|0\rangle$, and the rotation angle corresponds to the intrinsic probability according to equation \ref{angle}. For example, the $U3$ gate implementing the intrinsic probability on ${\rm RI}_3$, represented by qubit $q_1$, has an angle of $\theta=0.643$, corresponding to $P=10\%$.
The second rotation is executed if the controlling qubit is in state $|1\rangle$, and the rotation angle corresponds to the probability of being triggered by either the intrinsic probability or the transition probability. To illustrate this, consider ${\rm RI}_3$ again. If the controlling ${\rm RI}_2$ is triggered, then in 50\% of the cases, ${\rm RI}_3$ is triggered by the transition probability, regardless of the intrinsic outcome of ${\rm RI}_3$. In the remaining cases, it is still triggered intrinsically in 10\% of the cases. So if ${\rm RI}_2$ is on, then the probability of ${\rm RI}_3$ being triggered is 55\%. According to equation \ref{angle} this corresponds to $\theta = 1.671$. This concludes our description of how to implement the risk model as a quantum program. Compared to the implementation on a classical computer, the implementation of the model seems remarkably efficient: We use just $4$ qubits, $6$ elementary gates, and $4$ measurements of single qubits for one Monte Carlo simulation realization, including the random number generation!\\
Now we add control qubits that modify the probabilities for the risk items and transitions when switched on. We use these modifications to find the parameters that cause the biggest changes in our model. The following table shows which parameter is modified by which bit configuration for of those three bits:

\begin{table}[h!]
\begin{center}
\begin{tabular}{|l|c|c|c|}
$q_0$ & $q_1$ & $q_2$ & modification\\
\hline
0 & 0 & 0 & no mod.\\
0 & 0 & 1 & $p_1 +0.1$\\
0 & 1 & 0 & $p_3 +0.1$\\
0 & 1 & 1 & $p_4 +0.1$\\
1 & 0 & 0 & $p_{13} +0.1$\\
1 & 0 & 1 & $p_{14}+0.1$
\end{tabular}
\end{center}
\end{table}

For instance, this means that when $(q_0,q_1,q_2)=(0,1,0)$ the intrinsic probability $p_3$ of ${\rm RI}_3$ is increased by $0.1$. In our example, each modification increases one of the probabilities by $0.1$. The circuit diagram is shown in figure~\ref{modmodel}.

\hspace{0.5cm}
\begin{figure}[h]
\centering
\captionsetup{font=small,width=0.8\textwidth}
\scalebox{0.5}
{
\Qcircuit @C=1.0em @R=0.0em @!R {
\lstick{{q}_{0}:}&\qw&\ctrlo{1}&\qw&\qw&\qw&\ctrlo{1}&\qw&\qw&\ctrlo{1}&\ctrl{1}&\qw&\qw&\ctrlo{1}&\qw&\qw&\ctrlo{1}&\ctrl{1}&\qw&\qw\\
\lstick{{q}_{1}:}&\qw&\ctrlo{1}&\qw&\qw&\qw&\ctrl{1}&\qw&\qw&\ctrl{1}&\ctrlo{1}&\qw&\qw&\ctrl{1}&\qw&\qw&\ctrl{1}&\ctrlo{1}&\qw&\qw\\
\lstick{{q}_{2}:}&\qw&\ctrl{4}&\qw&\qw&\qw&\ctrlo{2}&\qw&\qw&\ctrlo{2}&\ctrlo{2}&\qw&\qw&\ctrl{1}&\qw&\qw&\ctrl{1}&\ctrl{1}&\qw&\qw\\
\lstick{{q}_{3}:}&\qw&\qw&\qw&\qw&\qw&\qw&\qw&\qw&\qw&\qw&\qw&\gate{U3(0.451)}&\gate{U3}&\qw&\gate{U3(1.430)}&\gate{U3}&\gate{U3}&\qw&\qw\\
\lstick{{q}_{4}:}&\qw&\qw&\qw&\qw&\gate{U3(0.643)}&\gate{U3}&\qw&\gate{U3(1.671)}&\gate{U3}&\gate{U3}&\qw&\qw&\qw&\qw&\qw&\qw&\qw&\qw&\qw\\
\lstick{{q}_{5}:}&\qw             &\qw      &\qw&\targ     &\ctrl{-1}&\ctrl{-1}&\qw&\ctrlo{-1}&\ctrl{-1}&\ctrl{-1}&\qw&\ctrlo{-2}&\ctrlo{-2}&\qw&\ctrl{-2}&\ctrl{-2}&\ctrl{-2}&\qw&\qw\\
\lstick{{q}_{6}:}&\gate{U3(2.214)}&\gate{U3}&\qw&\ctrlo{-1}&\qw      &\qw&\qw&\qw&\qw&\qw&\qw&\qw&\qw&\qw&\qw&\qw&\qw&\qw&\qw\\}
}
\caption{The quantum circuit diagram for the simplified risk model described in section \ref{usecase} with 4 risk items. When switched on, qubits $q_0$ to $q_2$ modify the probabilities with which the RI and transitions are triggered. A probability that corresponds to an intrinsic probability can only be modified by a single setting of the control qubits. A probability that is the result of an intrinsic and a transition probability can be modified by two settings of the control qubits.}
\label{modmodel}
\end{figure}
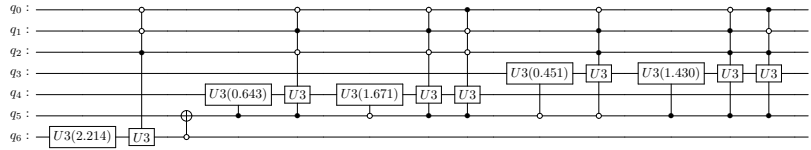

However, simply running this circuit and measuring the results repeatedly corresponds to a classical Monte Carlo simulation and scales just in the same way as a classical Monte Carlo simulation.\\
To gain a quantum advantage in the Monte Carlo simulation itself, we want to use this circuit in a QAE circuit~\cite{QAE}. First we show how to extend the circuit to account for the costs generated by each scenario:

\begin{figure}[h]
\centering
\captionsetup{font=small,width=0.8\textwidth}
\scalebox{1.1}
{
\Qcircuit @C=1.0em @R=0.2em @!R 
{
\lstick{ {q}_{0} :  } & \multigate{6}{{\rm risk\, model}} & \qw & \qw        & \qw      & \qw        & \qw      & \qw      & \qw       & \qw & \qw\\
\lstick{ {q}_{1} :  } & \ghost{riskmodel} & \qw      & \qw      & \qw        & \qw      & \qw        & \qw      & \qw      & \qw       & \qw\\
\lstick{ {q}_{2} :  } & \ghost{riskmodel} & \qw      & \qw      & \qw        & \qw      & \qw        & \qw      & \qw      & \qw       & \qw\\
\lstick{ {q}_{3} :  } & \ghost{riskmodel} & \qw      & \qw      & \qw        & \qw      & \qw        & \qw      & \ctrl{7} & \qw       & \qw\\
\lstick{ {q}_{4} :  } & \ghost{riskmodel} & \qw      & \qw      & \qw        & \qw      & \ctrl{5}   & \ctrl{5} & \qw      & \qw       & \qw\\
\lstick{ {q}_{5} :  } & \ghost{riskmodel} & \ctrl{2} & \ctrl{2} & \ctrl{2}   & \ctrl{2} & \qw        & \qw      & \qw      & \qw       & \qw\\
\lstick{ {q}_{6} :  } & \ghost{riskmodel} & \qw      & \qw      & \qw        & \qw      & \qw        & \qw      & \qw      & \qw       & \qw\\
\lstick{ {q}_{7} :  } & \qw               & \ctrl{1} & \ctrl{1} & \ctrl{1}   & \targ    & \qw        & \qw      & \qw      & \qw       & \qw\\
\lstick{ {q}_{8} :  } & \qw               & \ctrl{1} & \ctrl{1} & \targ      & \qw      & \qw        & \qw      & \qw      & \qw       & \qw\\
\lstick{ {q}_{9} :  } & \qw               & \ctrl{1} & \targ    & \qw        & \qw      & \ctrl{1}   & \targ    & \qw      & \ctrl{1}  & \qw\\
\lstick{ {q}_{10} :  }& \qw               & \targ    & \qw      & \qw        & \qw      & \targ      & \qw      & \targ    & \ctrl{1}  & \qw\\
\lstick{ {q}_{11} :  }& \qw               & \qw      & \qw      & \qw        & \qw      & \qw        & \qw      & \qw      & \targ     & \qw\\ 
}
}
\caption{The risk model with added cost register and indicator qubit ($q_{11}$) which indicates that the acceptable impact was exceeded.} \label{rmwithcosts}
\end{figure}
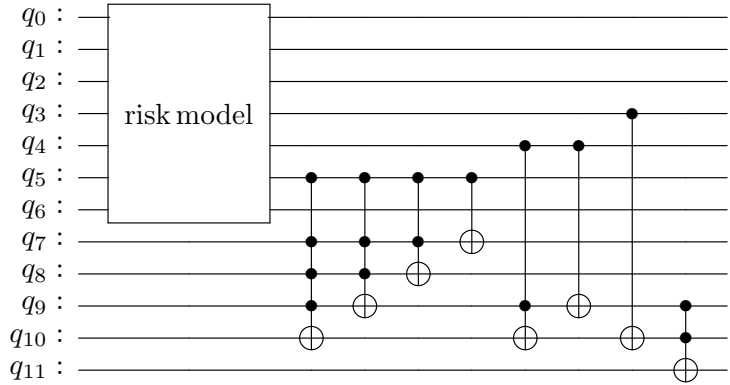

We have added qubits $q_7 ,q_8,q_9$ and $q_{10}$, which represent costs of 1,2,4 and 8 units, respectively. For instance, if $q_5$ is in state $|1\rangle$, then this means that ${\rm RI}_2$ is triggered and we have to add the value $1$ on this
register. This is done by the gates that are controlled by $q_5$. These gates implement an increment by $1$. The gates that are controlled by $q_4$ and $q_5$ implement an increment by $4$ and $8$, respectively. We omit an treatment of an overflow as this might not happen in our case.
In general, when qubit $q_i$ corresponds to a RI with costs $c_i$, then we add a circuit for adding $c_i$, which is controlled by $q_i$. If we write $c_i=\sum_k2^k c_i^{(k)}$ in binary representation then we could just add controlled increments for all values $k$ with $c_i^{(k)}=1$ to obtain the necessary total increment of $c_i$.  Finally, qubit $q_{11}$ indicates whether the acceptable impact of 12 units is exceeded or not. For our simple example, this is the case exactly when $q_{9}$ and $q_{10}$ are both switched on. This case is encoded by the final doubly controlled X gate in the circuit. In general, we need a circuit that compares the sum to the limit, that is set from outside. This can be done by adding the two complement of the limit to the calculated costs and checking the sign of the result. 

\subsection{The quantum risk model in QAE}
\label{quantumimplementation}
Now we are ready to use the risk model in a QAE circuit. The result that we can measure at the end of the circuit is the fraction of all possible outcomes which end up exceeding the acceptable financial impact, which we set to 12 units in our example. So this is a quantum way to do a Monte Carlo simulation of the risk model.\\
The key ingredient in the QAE circuit in figure~\ref{modelinQ} is the Grover operator built from the risk model. In~\cite{QAE} this operator is called $Q$ and for our case we denote it by ${\rm QRM}$. It is defined by 
$$
{\rm QRM} = - {\rm RM}\cdot S_0 \cdot {\rm RM^\dagger}\cdot S_X
$$ 
where ${\rm RM}$ is the unitary operator corresponding to the risk model circuit. We construct this circuit with an additional qubit $q_{12}$ for creating the necessary phases, e.g., the first controlled phase gate is applied if $q_{11}$ is activated. This operation corresponds to $S_X$.

\begin{figure}[h]
\centering
\captionsetup{font=small,width=0.8\textwidth}
\scalebox{0.7}
{
\Qcircuit @C=1.0em @R=0.0em @!R {    
\lstick{ {q}_{0} :  } & \qw& \qw & \multigate{11}{{\rm RM}^\dagger}& \qw & \qw & \qw & \qw & \qw & \qw & \multigate{11}{{\rm RM}} & \qw & \qw\\    
\lstick{ {q}_{1} :  } & \qw& \qw & \ghost{{\rm RM}^\dagger}& \qw & \qw & \qw & \qw & \qw & \qw & \ghost{{\rm RM}} & \qw & \qw\\    
\lstick{ {q}_{2} :  } & \qw& \qw & \ghost{{\rm RM}^\dagger}& \qw & \qw & \qw & \qw & \qw & \qw & \ghost{{\rm RM}} & \qw & \qw\\    
\lstick{ {q}_{3} :  } & \qw& \qw & \ghost{{\rm RM}^\dagger}& \qw & \ctrlo{1} & \gate{X} & \gate{Z} & \gate{X} & \gate{Z} & \ghost{{\rm RM}} & \qw & \qw\\
\lstick{ {q}_{4} :  } & \qw& \qw & \ghost{{\rm RM}^\dagger}& \qw & \ctrlo{1} & \qw & \qw & \qw & \qw & \ghost{{\rm RM}} & \qw & \qw\\    
\lstick{ {q}_{5} :  } & \qw& \qw & \ghost{{\rm RM}^\dagger}& \qw & \ctrlo{1} & \qw & \qw & \qw & \qw & \ghost{{\rm RM}} & \qw & \qw\\    
\lstick{ {q}_{6} :  } & \qw& \qw & \ghost{{\rm RM}^\dagger}& \qw & \ctrlo{1} & \qw & \qw & \qw & \qw & \ghost{{\rm RM}} & \qw & \qw\\    
\lstick{ {q}_{7} :  } & \qw& \qw & \ghost{{\rm RM}^\dagger}& \qw & \ctrlo{1} & \qw & \qw & \qw & \qw & \ghost{{\rm RM}} & \qw & \qw\\    
\lstick{ {q}_{8} :  } & \qw& \qw & \ghost{{\rm RM}^\dagger}& \qw & \ctrlo{1} & \qw & \qw & \qw & \qw & \ghost{{\rm RM}} & \qw & \qw\\    
\lstick{ {q}_{9} :  } & \qw& \qw & \ghost{{\rm RM}^\dagger}& \qw & \ctrlo{1} & \qw & \qw & \qw & \qw & \ghost{{\rm RM}} & \qw & \qw\\    
\lstick{ {q}_{10} :  } & \qw& \qw & \ghost{{\rm RM}^\dagger}& \qw & \ctrlo{1} & \qw & \qw & \qw & \qw & \ghost{{\rm RM}} & \qw & \qw\\    
\lstick{ {q}_{11} :  } & \qw& \ctrl{1} & \ghost{{\rm RM}^\dagger}& \qw & \ctrlo{1} & \qw & \qw & \qw & \qw & \ghost{{\rm RM}} & \qw & \qw\\    
\lstick{ {q}_{12} :  } & \qw& \ctrl{-1} &\qw & \gate{X} & \control\qw & \gate{X} & \qw & \qw & \qw & \qw & \qw & \qw\\  
}
}
\caption{This circuit implements the operator ${\rm QRM}$. This is the Grover operator using the risk model circuit ${\rm RM}$ described in section~\ref{themodel}. The controlled operators are phase gates $Z$. The factor $-1$ of the definition of ${\rm QRM}$ is included in the operations between ${\rm RM}^\dagger$ and ${\rm RM}$.}\label{modelinQ}
\end{figure}
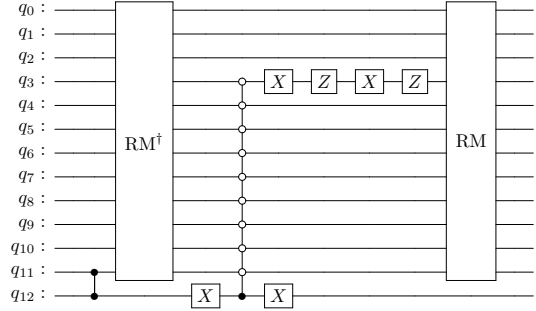

Next, we need to choose the resolution of the QAE with which we want to read out the answer. For our example, 8 qubits are acceptable. So we build the circuit of figure~\ref{modelinQAE} in which qubits $q_0$ to $q_7$ will carry the output and $q_8$ to $q_{10}$ carry the qubits which encode which input is modified, i.e., which of the probabilities of risk items and of transitions is increased. In the next section, these three qubits will make up the search space for the Grover search. The qubits $q_{11}$ to $q_{14}$ are the qubits corresponding to the risk items. The qubits $q_{15}$ to $q_{18}$ are the register for the costs. Qubit $q_{19}$ is used as the indicator bit which is marked when the Grover oracle detects a state meeting the search criteria (i.e., the costs are higher than $12$). Finally, qubit $q_{20}$ is added as an ancilla qubit to make the phase operations simpler. 

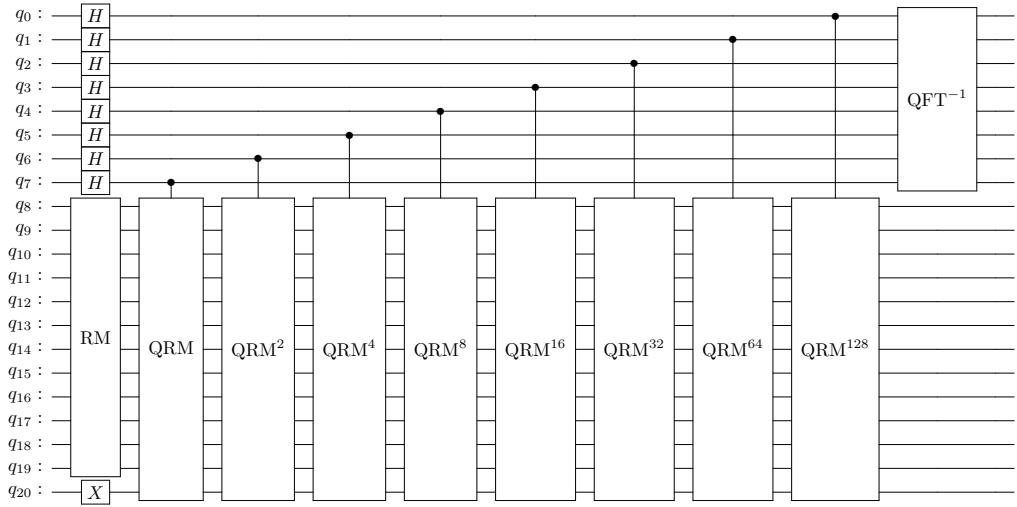
\begin{figure}[h]
\centering
\captionsetup{font=small,width=0.8\textwidth}
\scalebox{0.7}
{
\Qcircuit @C=1.0em @R=0.0em @!R
{
\lstick{ {q}_{0} :} & \gate{H} & \qw & \qw & \qw & \qw & \qw & \qw & \qw & \ctrl{8} & \multigate{7}{{\rm QFT}^{-1}} & \qw & \qw\\
\lstick{ {q}_{1} :} & \gate{H} & \qw & \qw & \qw & \qw & \qw & \qw & \ctrl{7} & \qw & \ghost{{\rm QFT}^{-1}} & \qw & \qw\\
\lstick{ {q}_{2} :} & \gate{H} & \qw & \qw & \qw & \qw & \qw & \ctrl{6} & \qw & \qw & \ghost{{\rm QFT}^{-1}} & \qw & \qw\\
\lstick{ {q}_{3} :} & \gate{H} & \qw & \qw & \qw & \qw & \ctrl{5} & \qw & \qw & \qw & \ghost{{\rm QFT}^{-1}} & \qw & \qw\\
\lstick{ {q}_{4} :} & \gate{H} & \qw & \qw & \qw & \ctrl{4} & \qw & \qw & \qw & \qw & \ghost{{\rm QFT}^{-1}} & \qw & \qw\\
\lstick{ {q}_{5} :} & \gate{H} & \qw & \qw & \ctrl{3} & \qw & \qw & \qw & \qw & \qw & \ghost{{\rm QFT}^{-1}} & \qw & \qw\\
\lstick{ {q}_{6} :} & \gate{H} & \qw & \ctrl{2} & \qw & \qw & \qw & \qw & \qw & \qw & \ghost{{\rm QFT}^{-1}} & \qw & \qw\\
\lstick{ {q}_{7} :} & \gate{H} & \ctrl{1} & \qw & \qw & \qw & \qw & \qw & \qw & \qw & \ghost{{\rm QFT}^{-1}} & \qw & \qw\\
\lstick{ {q}_{8} :} & \multigate{11}{\rm RM} & \multigate{12}{{\rm QRM}} & \multigate{12}{{\rm QRM}^2} & \multigate{12}{{\rm QRM}^4} & \multigate{12}{{\rm QRM}^8} & \multigate{12}{{\rm QRM}^{16}} & \multigate{12}{{\rm QRM}^{32}} & \multigate{12}{{\rm QRM}^{64}} & \multigate{12}{{\rm QRM}^{128}} & \qw & \qw & \qw\\
\lstick{ {q}_{9} :  } & \ghost{\rm RM} & \ghost{{\rm QRM}} & \ghost{{\rm QRM}^2} & \ghost{{\rm QRM}^4} & \ghost{{\rm QRM}^8} & \ghost{{\rm QRM}^{16}} & \ghost{{\rm QRM}^{32}} & \ghost{{\rm QRM}^{64}} & \ghost{{\rm QRM}^{128}} & \qw & \qw & \qw\\
\lstick{ {q}_{10} : } & \ghost{\rm RM} & \ghost{{\rm QRM}} & \ghost{{\rm QRM}^2} & \ghost{{\rm QRM}^4} & \ghost{{\rm QRM}^8} & \ghost{{\rm QRM}^{16}} & \ghost{{\rm QRM}^{32}} & \ghost{{\rm QRM}^{64}} & \ghost{{\rm QRM}^{128}} & \qw & \qw & \qw\\
\lstick{ {q}_{11} : } & \ghost{\rm RM} & \ghost{{\rm QRM}} & \ghost{{\rm QRM}^2} & \ghost{{\rm QRM}^4} & \ghost{{\rm QRM}^8} & \ghost{{\rm QRM}^{16}} & \ghost{{\rm QRM}^{32}} & \ghost{{\rm QRM}^{64}} & \ghost{{\rm QRM}^{128}} & \qw & \qw & \qw\\
\lstick{ {q}_{12} : } & \ghost{\rm RM} & \ghost{{\rm QRM}} & \ghost{{\rm QRM}^2} & \ghost{{\rm QRM}^4} & \ghost{{\rm QRM}^8} & \ghost{{\rm QRM}^{16}} & \ghost{{\rm QRM}^{32}} & \ghost{{\rm QRM}^{64}} & \ghost{{\rm QRM}^{128}} & \qw & \qw & \qw\\
\lstick{ {q}_{13} : } & \ghost{\rm RM} & \ghost{{\rm QRM}} & \ghost{{\rm QRM}^2} & \ghost{{\rm QRM}^4} & \ghost{{\rm QRM}^8} & \ghost{{\rm QRM}^{16}} & \ghost{{\rm QRM}^{32}} & \ghost{{\rm QRM}^{64}} & \ghost{{\rm QRM}^{128}} & \qw & \qw & \qw\\
\lstick{ {q}_{14} : } & \ghost{\rm RM} & \ghost{{\rm QRM}} & \ghost{{\rm QRM}^2} & \ghost{{\rm QRM}^4} & \ghost{{\rm QRM}^8} & \ghost{{\rm QRM}^{16}} & \ghost{{\rm QRM}^{32}} & \ghost{{\rm QRM}^{64}} & \ghost{{\rm QRM}^{128}} & \qw & \qw & \qw\\
\lstick{ {q}_{15} : } & \ghost{\rm RM} & \ghost{{\rm QRM}} & \ghost{{\rm QRM}^2} & \ghost{{\rm QRM}^4} & \ghost{{\rm QRM}^8} & \ghost{{\rm QRM}^{16}} & \ghost{{\rm QRM}^{32}} & \ghost{{\rm QRM}^{64}} & \ghost{{\rm QRM}^{128}} & \qw & \qw & \qw\\
\lstick{ {q}_{16} : } & \ghost{\rm RM} & \ghost{{\rm QRM}} & \ghost{{\rm QRM}^2} & \ghost{{\rm QRM}^4} & \ghost{{\rm QRM}^8} & \ghost{{\rm QRM}^{16}} & \ghost{{\rm QRM}^{32}} & \ghost{{\rm QRM}^{64}} & \ghost{{\rm QRM}^{128}} & \qw & \qw & \qw\\
\lstick{ {q}_{17} : } & \ghost{\rm RM} & \ghost{{\rm QRM}} & \ghost{{\rm QRM}^2} & \ghost{{\rm QRM}^4} & \ghost{{\rm QRM}^8} & \ghost{{\rm QRM}^{16}} & \ghost{{\rm QRM}^{32}} & \ghost{{\rm QRM}^{64}} & \ghost{{\rm QRM}^{128}} & \qw & \qw & \qw\\
\lstick{ {q}_{18} : } & \ghost{\rm RM} & \ghost{{\rm QRM}} & \ghost{{\rm QRM}^2} & \ghost{{\rm QRM}^4} & \ghost{{\rm QRM}^8} & \ghost{{\rm QRM}^{16}} & \ghost{{\rm QRM}^{32}} & \ghost{{\rm QRM}^{64}} & \ghost{{\rm QRM}^{128}} & \qw & \qw & \qw\\
\lstick{ {q}_{19} : } & \ghost{\rm RM} & \ghost{{\rm QRM}} & \ghost{{\rm QRM}^2} & \ghost{{\rm QRM}^4} & \ghost{{\rm QRM}^8} & \ghost{{\rm QRM}^{16}} & \ghost{{\rm QRM}^{32}} & \ghost{{\rm QRM}^{64}} & \ghost{{\rm QRM}^{128}} & \qw & \qw & \qw\\
\lstick{ {q}_{20} : } & \gate{X} & \ghost{{\rm QRM}} & \ghost{{\rm QRM}^2} & \ghost{{\rm QRM}^4} & \ghost{{\rm QRM}^8} & \ghost{{\rm QRM}^{16}} & \ghost{{\rm QRM}^{32}} & \ghost{{\rm QRM}^{64}} & \ghost{{\rm QRM}^{128}} & \qw & \qw & \qw\\
}
}
\caption{This circuit implements QAE on the risk model RM. The output can be measured with an $8$ bit precision on $q_0$ to $q_7$. RM is the risk model circuit as in figure~\ref{rmwithcosts}, QRM is the operator Q from \cite{QAE}, that is, the Grover operator using RM in the oracle as in figure \ref{modelinQ}, and ${\rm QFT}^{-1}$ is the inverse quantum Fourier transform.}
\label{modelinQAE}
\end{figure}

It is instructive to run this circuit and measure the output. With 100 shots, we obtain the results in
figure~\ref{QAEoutput}.
\begin{figure}[h]
\centering
\captionsetup{font=small,width=0.8\textwidth}
\includegraphics[width=85mm]{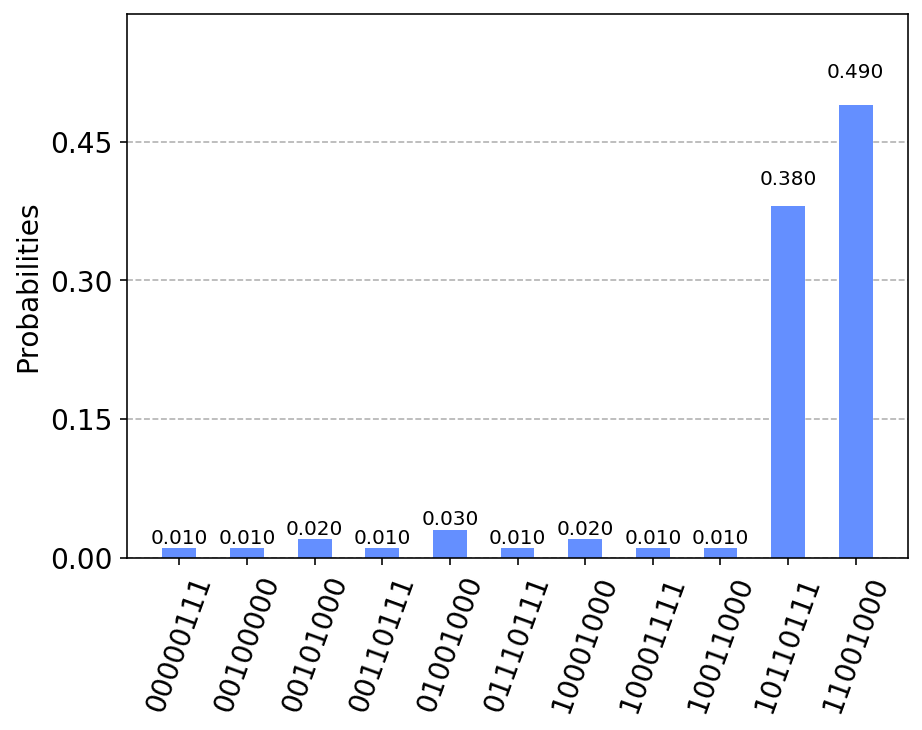}
\caption{The output of the QAE on the risk model with all input modifications (qubits $q_8$, $q_9$ and $q_{10}$) set to zero, i.e., the probabilities are not modified. The first digit in the binary strings is the least significant bit. This corresponds to the standard Monte Carlo analysis without sensitivity analysis. One could also run this analysis for each possible modification of the inputs, save the results and search the result for the biggest input on a classical computer. This would scale linearly in the number of modifications.}\label{QAEoutput}
\end{figure}
We see that the 100 runs of the quantum program delivered 11 different answers, but with 87 runs distributed between the two values 19 (11001000 in binary) and 237 (10110111 in binary) at the far right of the histogram. We remind the reader that the result we measure is a fraction in the interval [0,1] in binary encoding. That means that we find the answers $a=19/256$ and $a=237/256$. These two results are actually two different versions of the same answer, as can be seen when we convert the output $a$ of the QAE to the probability 
\begin{equation}
P(a) = \sin^2( a \pi)\;.
\end{equation}
For $a=19/256$ and $a=237/256$ we obtain $P(a) \approx 0.0534$.
To summarize, the question the QAE answers here is this: What is the probability that the risk model finds an impact of 12 or larger? The answer that our quantum program delivers is that 12 will be exceeded in 5.34\% of the cases, and this answer is found with 87\% probability. This means the circuit in figure \ref{modelinQAE} is a quantum implementation of the classical Monte Carlo simulation which is currently performed at Deutsche Börse Group to analyze the business risks. 

\subsection{The Grover search}\label{sectionGroverSearch}
The final step in the implementation is to use the circuit of figure~\ref{modelinQAE} that performs the QAE on the model in a Grover search. The search should find the modification settings (qubits $q_8$ to $q_{10}$) that lead to a certain probability that the loss is above the threshold. For instance, we could be interested in finding the modification settings that lead to a loss probability of $7.75\%$. This probability corresponds to the binary encodings $11101000$ of $23$ and $10010111$ of $233$ that are the interesting outputs of QAE. Since the probabilities are encoded in qubits $q_0$ to $q_7$ by the QAE, we use phase operations, which are controlled by these qubits, to construct a Grover oracle that searches a modification setting that leads to the desired probability of $7.75\%$.
The resulting circuit is given in figure \ref{grover}. The first application of the QAE leads to a superposition of states that lead to the results of figure~\ref{QAEoutput} when measured. The boxed part of the circuit is the oracle. This oracle is at the core of this paper: it is a unitary oracle based on QAE and it is an imperfect oracle. The oracle applies a phase operation on $q_{21}$ conditioned on the two solutions $11101000$ and $10010111$. This operation leads to the phase $-1$ for these two solutions and to the phase $1$ for all other solutions. After this phase operation, we apply the inverse of the QAE circuit. As a consequence, the phase $-1$ is applied for these two states that correspond to the probability ${\rm sin}(a\pi)^2$ with $a=23/256$ or $a=233/256$ that the threshold of loss (12 units in our example) is reached. The Grover search is looking for a configuration of $q_8$, $q_9$ and $q_{10}$ that leads to this probability. Strictly speaking, we should also check all solutions bigger than $7.75\%$ on top of that, but in order to unclutter the diagram and since we know that there are no such solutions, we left those out.
\begin{figure}[h]
\centering
\captionsetup{font=small,width=0.8\textwidth}
\scalebox{0.75}{
\Qcircuit @C=1.0em @R=0.0em @!R {
\lstick{ {q}_{0} :  } & \qw & \multigate{20}{{\rm model\,QAE}} & \ctrlo{1} & \ctrl{1} & \multigate{20}{{\rm model\,QAE}^\dagger} & \qw & \qw & \qw & \qw & \qw & \qw & \qw & \qw & \qw & \qw & \qw & \qw\\
\lstick{ {q}_{1} :  } & \qw & \ghost{{\rm model QAE}} & \ctrlo{1} & \ctrl{1} & \ghost{{\rm model\,QAE}^\dagger} & \qw & \qw & \qw & \qw & \qw & \qw & \qw & \qw & \qw & \qw & \qw & \qw\\
\lstick{ {q}_{2} :  } & \qw & \ghost{{\rm model QAE}} & \ctrlo{1} & \ctrl{1} & \ghost{{\rm model\,QAE}^\dagger} & \qw & \qw & \qw & \qw & \qw & \qw & \qw & \qw & \qw & \qw & \qw & \qw\\
\lstick{ {q}_{3} :  } & \qw & \ghost{{\rm model QAE}} & \ctrl{1} & \ctrlo{1} & \ghost{{\rm model\,QAE}^\dagger} & \qw & \qw & \qw & \qw & \qw & \qw & \qw & \qw & \qw & \qw & \qw & \qw\\
\lstick{ {q}_{4} :  } & \qw & \ghost{{\rm model QAE}} & \ctrlo{1} & \ctrl{1} & \ghost{{\rm model\,QAE}^\dagger} & \qw & \qw & \qw & \qw & \qw & \qw & \qw & \qw & \qw & \qw & \qw & \qw\\
\lstick{ {q}_{5} :  } & \qw & \ghost{{\rm model QAE}} & \ctrl{1} & \ctrlo{1} & \ghost{{\rm model\,QAE}^\dagger} & \qw & \qw & \qw & \qw & \qw & \qw & \qw & \qw & \qw & \qw & \qw & \qw\\
\lstick{ {q}_{6} :  } & \qw & \ghost{{\rm model QAE}} & \ctrl{1} & \ctrlo{1} & \ghost{{\rm model\,QAE}^\dagger} & \qw & \qw & \qw & \qw & \qw & \qw & \qw & \qw & \qw & \qw & \qw & \qw\\
\lstick{ {q}_{7} :  } & \qw & \ghost{{\rm model QAE}} & \ctrl{14} & \ctrl{14} & \ghost{{\rm model\,QAE}^\dagger} & \qw & \qw & \qw & \qw & \qw & \qw & \qw & \qw & \qw & \qw & \qw & \qw\\
\lstick{ {q}_{8} :  } & \gate{H} & \ghost{{\rm model QAE}} & \qw & \qw & \ghost{{\rm model\,QAE}^\dagger} & \gate{H} & \gate{X} & \qw & \ctrl{1} & \gate{X} & \gate{H} & \qw & \meter & \qw & \qw & \qw & \qw\\
\lstick{ {q}_{9} :  } & \gate{H} & \ghost{{\rm model QAE}} & \qw & \qw & \ghost{{\rm model\,QAE}^\dagger} & \gate{H} & \gate{X} & \qw & \ctrl{1} & \gate{X} & \gate{H} & \qw & \qw & \meter & \qw & \qw & \qw\\
\lstick{ {q}_{10} :  } & \gate{H} & \ghost{{\rm model QAE}} & \qw & \qw & \ghost{{\rm model\,QAE}^\dagger} & \gate{H} & \gate{X} & \gate{H} & \targ & \gate{H} & \gate{X} & \gate{H} & \qw & \qw & \meter & \qw & \qw\\
\lstick{ {q}_{11} :  } & \qw & \ghost{{\rm model QAE}} & \qw & \qw & \ghost{{\rm model\,QAE}^\dagger} & \qw & \qw & \qw & \qw & \qw & \qw & \qw & \qw & \qw & \qw & \qw & \qw\\
\lstick{ {q}_{12} :  } & \qw & \ghost{{\rm model QAE}} & \qw & \qw & \ghost{{\rm model\,QAE}^\dagger} & \qw & \qw & \qw & \qw & \qw & \qw & \qw & \qw & \qw & \qw & \qw & \qw\\
\lstick{ {q}_{13} :  } & \qw & \ghost{{\rm model QAE}} & \qw & \qw & \ghost{{\rm model\,QAE}^\dagger} & \qw & \qw & \qw & \qw & \qw & \qw & \qw & \qw & \qw & \qw & \qw & \qw\\
\lstick{ {q}_{14} :  } & \qw & \ghost{{\rm model QAE}} & \qw & \qw & \ghost{{\rm model\,QAE}^\dagger} & \qw & \qw & \qw & \qw & \qw & \qw & \qw & \qw & \qw & \qw & \qw & \qw\\
\lstick{ {q}_{15} :  } & \qw & \ghost{{\rm model QAE}} & \qw & \qw & \ghost{{\rm model\,QAE}^\dagger} & \qw & \qw & \qw & \qw & \qw & \qw & \qw & \qw & \qw & \qw & \qw & \qw\\
\lstick{ {q}_{16} :  } & \qw & \ghost{{\rm model QAE}} & \qw & \qw & \ghost{{\rm model\,QAE}^\dagger} & \qw & \qw & \qw & \qw & \qw & \qw & \qw & \qw & \qw & \qw & \qw & \qw\\
\lstick{ {q}_{17} :  } & \qw & \ghost{{\rm model QAE}} & \qw & \qw & \ghost{{\rm model\,QAE}^\dagger} & \qw & \qw & \qw & \qw & \qw & \qw & \qw & \qw & \qw & \qw & \qw & \qw\\
\lstick{ {q}_{18} :  } & \qw & \ghost{{\rm model QAE}} & \qw & \qw & \ghost{{\rm model\,QAE}^\dagger} & \qw & \qw & \qw & \qw & \qw & \qw & \qw & \qw & \qw & \qw & \qw & \qw\\
\lstick{ {q}_{19} :  } & \qw & \ghost{{\rm model QAE}} & \qw & \qw & \ghost{{\rm model\,QAE}^\dagger} & \qw & \qw & \qw & \qw & \qw & \qw & \qw & \qw & \qw & \qw & \qw & \qw\\
\lstick{ {q}_{20} :  } & \qw & \ghost{{\rm model QAE}} & \qw & \qw & \ghost{{\rm model\,QAE}^\dagger} & \qw & \qw & \qw & \qw & \qw & \qw & \qw & \qw & \qw & \qw & \qw & \qw\\
\lstick{ {q}_{21} :  } & \gate{X} & \qw & \gate{Z} & \gate{Z} & \qw & \qw & \qw & \qw & \qw & \qw & \qw & \qw & \qw & \qw & \qw & \qw & \qw\\
\lstick{c:}            & {/_{_{3}}} \cw & \cw & \cw & \cw & \cw & \cw & \cw & \cw & \cw & \cw & \cw & \cw & \dstick{0} \cw \cwx[-14] & \dstick{1} \cw \cwx[-13] & \dstick{2} \cw \cwx[-12] & \cw & \cw \gategroup{1}{3}{21}{6}{.5em}{--}
}}
\caption{The quantum circuit which implements the sensitivity analysis for the risk model with one Grover step. The qubits $q_0$ to $q_7$ are the outputs of the QAE where $q_0$ the least significant bit of the binary representation. The Grover oracle consists of the QAE, the two conditioned phase operations and the inverse QAE. The boxed part of the circuit is the imperfect, QAE-based oracle.}\label{grover}
\end{figure}
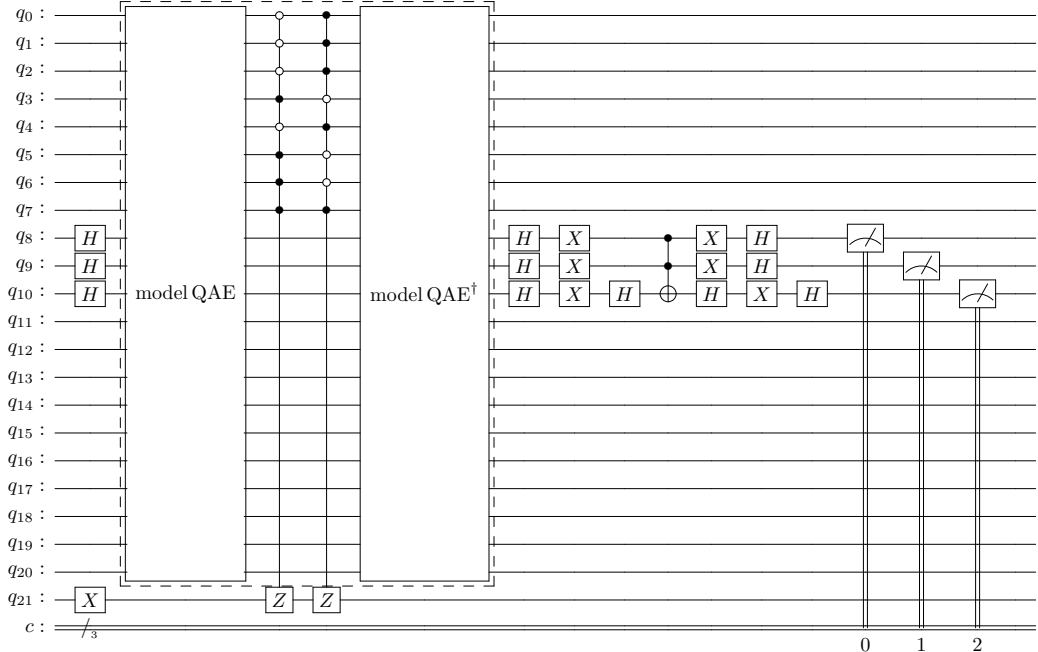
The result of our quantum program, which consists of one Grover step, leads to the result histogram of figure~\ref{figureHistogram}.
\begin{figure}[h]
\centering
\captionsetup{font=small,width=0.8\textwidth}
\includegraphics[width=85mm]{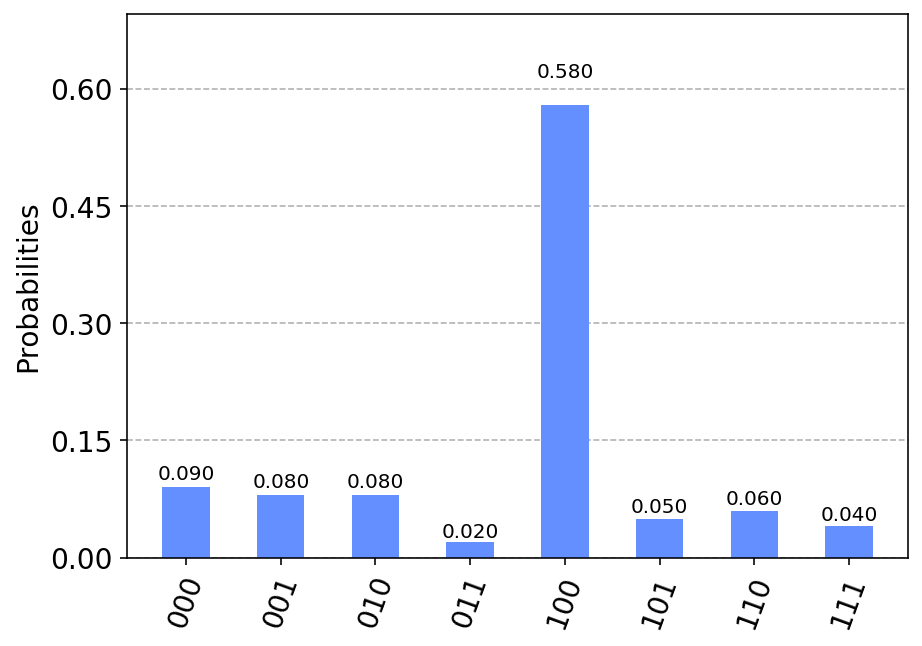}
\caption{The output of the sensitivity analysis. The first digit of a binary string is the least significant bit. The solution encoded by 100, corresponding to risk item 1, is correctly identified in 58\% of the cases.}
\label{figureHistogram}
\end{figure}
This means that a single Grover step leads to the solution $001$ with 58\% probability. This corresponds to the modification of ${\rm RI}_1$ in our example (see section \ref{usecase}). The success probability is lower than the 81\% one might expect from the QAE. The reason is that in a search space of just 3 bits the optimal number of search steps is not exactly 1 and we cannot perform fractional steps. Therefore, we cannot get closer to the maximal probability. The same is true for the search with standard Grover oracles when the search space is small. When the search space is large, we typically do better than $81\%$.

\section{Scaling experiments with the quantum model}\label{scalingexperiments}
In section \ref{quantum} we showed how to implement the sensitivity analysis of the risk model for a toy version with just four risk items. However, the model used in production uses several hundred risk items. Therefore, the key question is how the performance of the quantum model scales with the size of the model. In this section, we will shed some light on this question by simulating risk models of different sizes and comparing their complexity.\\
Since the memory requirements of simulations of quantum computers grow exponentially in the number of simulated qubits, we had to restrict the simulations to toy versions of the actual risk model with up to about $20$ qubits in total.\\
To test the scaling under somewhat realistic conditions, we set up a risk model with $7$ risk items and $6$ transitions and performed the sensitivity analysis on it. The costs for each risk item were set to $1$, thus allowing us to cut off the register representing the costs as in figure~\ref{rmwithcosts}. Unfortunately, trials with a more complex cost structure pushed the simulator beyond the limits of our computer with 250GB of RAM.

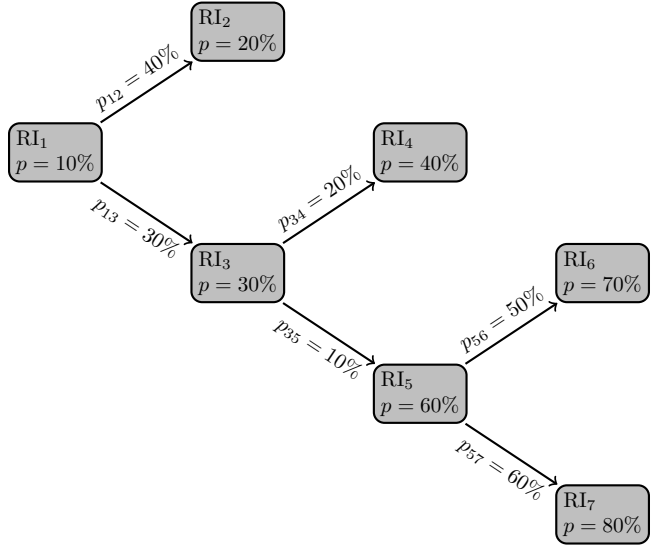
\begin{figure}[h]
\centering
\captionsetup{font=small,width=0.8\textwidth}
\begin{tikzpicture}[thick,scale=0.8, every node/.style={transform shape}]
\node[rectangle,rounded corners,draw=black,align=left,fill=lightgray] (one) at (-2,0) {${\rm RI}_1$ \\$p=10\%$};
\node[rectangle,rounded corners,draw=black,align=left,fill=lightgray] (two) at (1, 2) {${\rm RI}_2$ \\$p=20\%$};
\node[rectangle,rounded corners,draw=black,align=left,fill=lightgray] (three) at (1, -2) {${\rm RI}_3$ \\$p=30\%$};
\node[rectangle,rounded corners,draw=black,align=left,fill=lightgray] (four) at (4, 0) {${\rm RI}_4$\\ $p=40\%$};
\node[rectangle,rounded corners,draw=black,align=left,fill=lightgray] (five) at (4, -4) {${\rm RI}_5$\\ $p=60\%$};
\node[rectangle,rounded corners,draw=black,align=left,fill=lightgray] (six) at (7, -2) {${\rm RI}_6$\\ $p=70\%$};
\node[rectangle,rounded corners,draw=black,align=left,fill=lightgray] (seven) at (7, -6) {${\rm RI}_7$\\ $p=80\%$};
\path[->,draw,thick] (one) edge node[sloped, above] {$p_{12}=40\%$} (two);
\path[->,draw,thick] (one) edge node[sloped, below] {$p_{13}=30\%$} (three);
\path[->,draw,thick] (three) edge node[sloped, above] {$p_{34}=20\%$} (four);
\path[->,draw,thick] (three) edge node[sloped, below] {$p_{35}=10\%$} (five);
\path[->,draw,thick] (five) edge node[sloped, above] {$p_{56}=50\%$} (six);
\path[->,draw,thick] (five) edge node[sloped, below] {$p_{57}=60\%$} (seven);
\end{tikzpicture}
\caption{The structure of the risk model with $7$ risk items which we used for the scaling experiments. Each node represents a risk item. The figure shows the model with $7$ RI. In the experiment with $6$ risk items, we cut off ${\rm RI}_7$. In the experiment with $5$ RI, we cut off ${\rm RI}_7$ and ${\rm RI}_6$, and similarly for $4$ and $3$ RI. The smallest model in the test consisted of just $2$ RI, with just ${\rm RI}_1$ and ${\rm RI}_2$ in the figure.}\label{scalingtree}
\end{figure}

As described earlier, the question we want to answer with the quantum algorithm is the
following: Is there a parameter which, when changed slightly, pushes us across a predefined acceptable limit for the probability of extreme cases? For our experiments with $N$ risk items and with the simplified toy model, we set the maximal acceptable cash impact to $N$ units, i.e., we have a critical event only when all risk items are triggered.
We set the acceptable probability for each case separately, so that there was one single risk item, which had a much higher impact compared to the others. The task of the quantum algorithm was to find out which one it was in each case.\\
If applied in practice, one would have to run the simulation several times. If each run gives different solutions, then this would indicate that there is no single parameter which dominates the sensitivity.
Note that the number of Grover steps in the circuit increases when the number of RI increases. For $2$, $3$ and $4$ risk items, one Grover step is required, for $5$, $6$ and $7$ risk items two Grover steps are required. The estimate of how many Grover steps are optimal is discussed in the theory section~\ref{theory}.\\
Note that the classical Monte Carlo simulation that we used for result comparisons calculates more information than our quantum algorithm. For example, we get a ranking of the impact of all parameters with little extra effort, whereas the quantum algorithm will only find the parameter with the biggest impact.\\
For the described setup, we required the classical Monte Carlo simulation and the quantum program to identify the parameter (risk item or transition probability) which pushes the cash impact beyond the acceptable limit (i.e., all risk items were triggered) with at least $70\%$ probability. We then compared how many Monte Carlo runs were required to achieve this on the classical side, and how many times the model code is executed on the quantum side. Note that this is twice the number of calls to the QAE oracle by our construction. The results of our numerical tests are displayed in figure~\ref{scaling}.
\begin{figure}[h]
\centering
\captionsetup{font=small,width=0.8\textwidth}
\includegraphics[width=80mm]{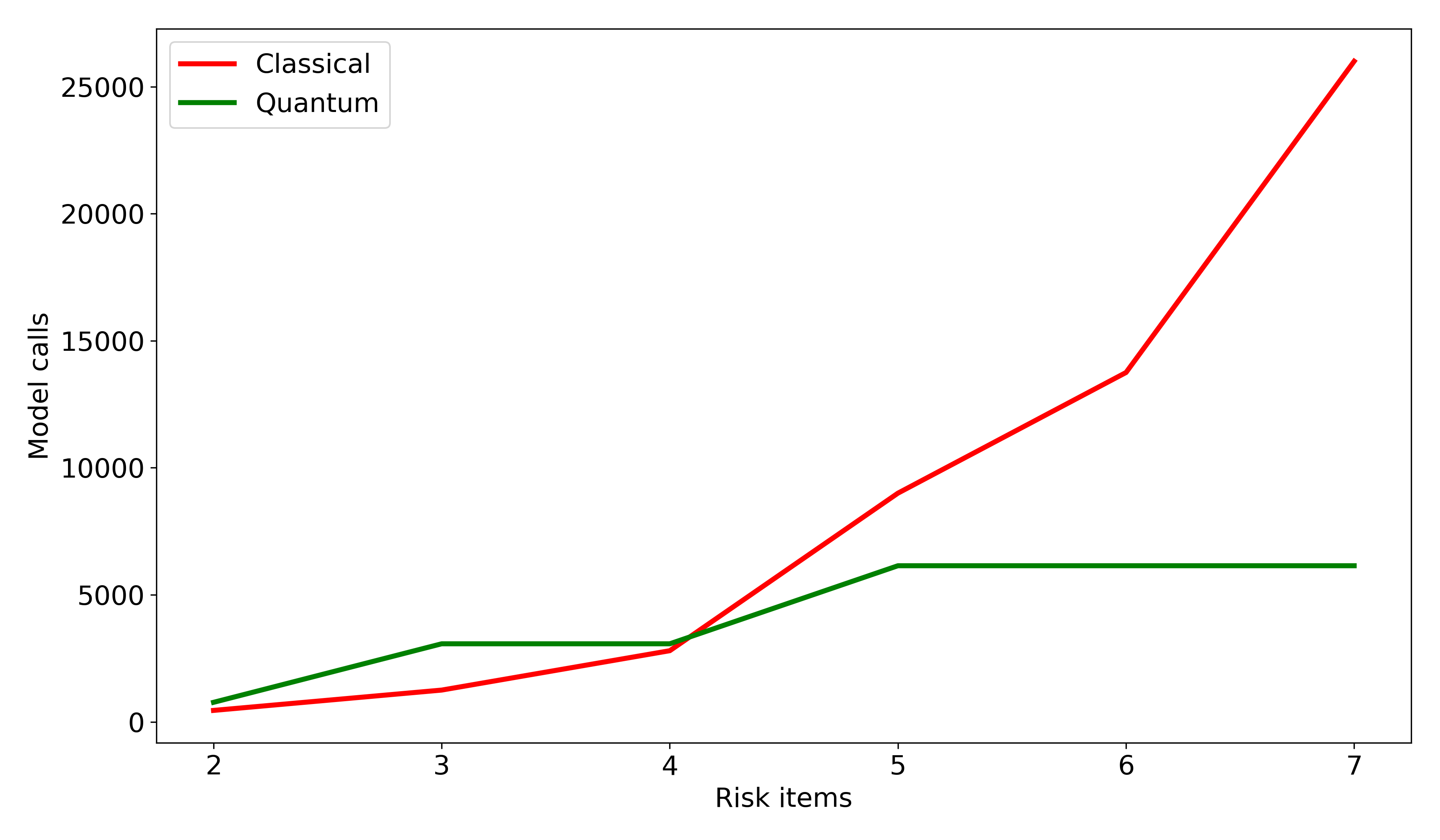}
\caption{The scaling of the quantum program versus classical Monte Carlo simulation. The objective was to identify the parameter with the biggest impact on breach probability with at least 70\% probability. On the y-axis, we see how often the risk model had to be evaluated (i.e., Monte Carlo runs for the classical cases and calls to the risk model in the quantum code).}\label{scaling}
\end{figure}
The scaling behavior of the classical results is a mixture of quadratic growth in the precision required from the Monte Carlo simulation (but not directly in the number of risk items, even though larger models require higher precision), and a linear growth in the number of risk items from the search of the parameter with the biggest impact. The scaling of the quantum results is linear in the required precision and the square root of the number of risk items. So the quantum program should scale quadratically better than the classical solution, which seems in line with our results in figure~\ref{scaling}.\\
Our scaling test is a simplification of the actual application in several ways: First of all, we assigned equal costs to all events, which simplifies the problem. Without that simplification, the required computing resources for the simulation would have exceeded the capabilities of our computers due to the extra qubits for the cost register. In section~\ref{themodel} we described a quantum circuit that deals efficiently with costs, so this will not be an issue once appropriate quantum hardware is available. Second, we used the fact there was exactly one parameter which would breach the limit. In practice, there might be none (which would show in the quantum program by random results), or several (which would not be an issue). Third, a combination of two, three or more parameter modifications might have a very large impact, whereas each of them separately might not. While trying all possible tuples is a problem that grows exponentially in the number of parameters and thus is out of reach even of the quantum program, in practice we would expect it to be sufficient to restrict the analysis to pairs. For instance, it is easy to imagine that the estimation of one parameter might be off and that this turns out to have a large impact, whereas being off in three or more parameters with a large impact seems to be a much lower risk. Furthermore, practical experience and some numerical experiments suggest that the model is relatively robust against perturbations.\\
The results seem to indicate that our quantum implementation does indeed work and that it scales as intended, thus giving a quadratic improvement in complexity over the search through the results of the classical Monte Carlo solutions.

\section{Outlook: Requirements for use in production}\label{outlook}
From a business point of view, our results regarding the scaling of the quantum program look promising, but the key question is: What would be required in terms of actual quantum hardware to run the full model? In this section, we make rough estimates regarding the number of (error corrected) qubits such a machine would require, and how many (error corrected) gates the code would require.

\subsection{Number of qubits}
Estimating the number of required qubits is straightforward. We saw in section~\ref{themodel} that our implementation needs one qubit for each of the risk items in the model, regardless of the details of the dependencies (transition probabilities) between them.
Next, we need a register for the modifications. We encoded the number of the modified risk items in binary, so if we have $n_r$ risk items and $n_t$ transition probabilities we need ${\rm log}_2(n_r+n_t)$ qubits for the controls. We denote the number of control qubits by $n_s$ and this constitutes the search space for the Grover search.\\
Furthermore, we need a register for the output of the amplitude estimation. We denote the size of this register by $n_{\rm ae}$. In practice, since one is always interested in tail events with probabilities close to $0$ and this is where amplitude estimation is most sensitive, we need a relatively small number of qubits. For example, with $n_{\rm ae}=10$ qubits precision, we already have more than $30$ possible values for the QAE in the range between $0\%$ and $1\%$.\\
We will also need a register for the costs. Since we also encode this in binary representation with $n_c$ qubits, we could encode costs from $0$ to $1023$ units with just $n_c=10$ qubits.\\
To pick an example of realistic magnitude, let's say we have $150$ risk items and $250$ transition probabilities. We want the most expensive risk to be $1000$ times more expensive than the cheapest. The percentage range we want to exclude would be the worst $0.01\%$. So according to our argument above we could work with $150 + 10 + 12 + 10 = 182$ qubits.\\
So we claim that with as few as roughly $200$ (error corrected) qubits, the algorithm could tackle problems of practical relevance, which are impractical to perform on classical hardware.

\subsection{Number of gates}
In this section we give an estimation for the number of gates for our quantum implementation of the sensitivity analysis. We assume that we have $n_r$ risk items and $n_t$ transitions.

We first consider the complexity of the quantum implementation of the risk model. 
For a risk item ${\rm RI}_i$, that is not triggered by a transition probability, we only need a $U3(\theta_i, 0, 0)$ gate on the corresponding qubit with $\theta_i=2 \cdot {\rm arcsin}(\sqrt{p_i})$ where $p_i$ denotes the intrinsic probability that ${\rm RI}_i$ is triggered. After applying this gate we have the amplitudes $\sqrt{1-p_i}$ for $|0\rangle$ and $\sqrt{p_i}$ for $|1\rangle$. As there are at most $n_r$ risk items that are not triggered by other risk items, we have $O(n_r)$ uncontrolled $U3$ gates.

For a risk item ${\rm RI}_j$ that is triggered by risk items ${\rm RI}_{i_0}, \ldots, {\rm RI}_{i_{m-1}}$ we could calculate the $2^m$ probabilities $p_\alpha$ for $\alpha=(\alpha_0, \ldots, \alpha_{m-1})$ with $\alpha_k=1$ if risk item ${\rm RI}_{i_k}$ is active and $\alpha_k=0$ else. We use the $2^m$ gates $U3(\theta_\alpha,0,0)$ with $\theta_\alpha=2 \cdot {\rm arcsin}(\sqrt{p_\alpha})$ that are controlled by the qubits corresponding to ${\rm RI}_{i_0}, \ldots, {\rm RI}_{i_{m-1}}$. If $\alpha_i=0$ then the corresponding qubit is used with a negative control. The number of $U3$ gates with this construction is $O(2^m)$ and each gate is controlled by $m$ qubits. Obviously, this is only efficient for a fixed number $m$. 

An alternative construction for risk items that are triggered by $m$ other risk items is the construction of a binary tree. For simplicity, assume that $m=2^k$ and then we consider the pairs $({\rm RI}_{i_0},{\rm RI}_{i_1}), \ldots, ({\rm RI}_{i_{m-2}},{\rm RI}_{i_{m-1}})$. For a pair $({\rm RI}_{i_\ell}, {\rm RI}_{i_{\ell+1}})$ we introduce an ancilla risk item with zero costs and we use the construction from above with $m=2$ to trigger this ancilla RI by ${\rm RI}_\ell$ and ${\rm RI}_{\ell+1}$ with the correct probabilities. Then in the next step, all ancilla risk items are grouped to pairs again and each pair leads to another new ancilla risk item, but now with probability $1$. This construction needs $k$ layers of combining two risk items each in order to lead to the risk item ${\rm RI}_i$. Since there are at most $n_t$ risk items with a transition, we need $O(n_t)$ ancilla qubits and $O(n_t)$ constructions of amplitudes from pairs of risk items. Each of this amplitude calculations needs $4$ doubly-controlled $U3$ operations by the previous construction.

We next consider the circuit corresponding to a risk item ${\rm RI}_i$ that triggers exclusively one of the risk items ${\rm RI}_{j_0}, \ldots, {\rm RI}_{j_{m-1}}$ with probabilities $p_0, \ldots, p_{m-1}$ that satisfy $\sum_i p_i=1$. We add the gate $U3(\theta_0, 0, 0)$ with $\theta_0=2 \cdot{\rm arcsin}(\sqrt{p_0})$ to the qubit for ${\rm RI}_{j_0}$. The $U3$ operation is triggered when the qubit for ${\rm RI}_i$ is $1$. Then we add the gate $U3(\theta^\prime_1,0,0)$ with $\theta^\prime = \theta_1/\theta_0$ to the qubit for ${\rm RI}_{j_1}$ that is triggered when the qubit for ${\rm RI}_{j_0}$ is $0$ and so on. Note, that in each step we have to successively calculate new transition probability $\theta_i^\prime$. This construction leads to the desired probabilities of the XOR cluster. Since the maximum number of transitions is $n_t$ the number of single-controlled $U3$ gates is $O(n_t)$. Note that if several risk items ${\rm RI}_{i_0}, \ldots, {\rm RI}_{i_{\ell-1}}$ trigger a XOR cluster ${\rm RI}_{j_0}, \ldots, {\rm RI}_{j_{m-1}}$ of risk items, then we can introduce an ancilla risk item ${\rm RI}_k$ with no costs and split the transition into two parts: The first part is the transition from the ${\rm RI}_i$ to ${\rm RI}_k$ and the second part is the transition from ${\rm RI}_k$ to the ${\rm RI}_j$.

In our circuits, we have an additional register of $n_c$ qubits for the modification of the intrinsic and transition probabilities. For a risk item a modification changes only a block of the circuit that calculates the amplitudes for the corresponding qubit. The parts before and after this block are not affected. Therefore, we can just add two new blocks directly behind the original block. Both blocks are only activated when the $n_c$ control qubits have the correct configuration. The first new block is the inverse of the original block and the second new block is the same but with modified values to respect the modifications of probabilities. Therefore, this construction leads $m$ gates that are controlled by $k$ qubits to $3m$ gates that are controlled by $n_c+k$ qubits. Note that $k$ is $O({\rm log}(n_r+n_t))$.

For the addition of the costs we consider the construction of figure~\ref{rmwithcosts}. The gates that are controlled by $q_5$ implement the operation $+1$ for a binary representation on $4$ qubits and the gates can be easily generalized to the operation $+1$ on $n_c$ qubits. For the addition $+1$ on $n_c$ qubits we need at most $n_c$ NOT gates that are controlled by at most $n_c-1$ qubits. If we add another qubit without any gate below this circuit then we obtain the operation $+2$. Therefore, we can easily construct circuits for adding any power of two. If a risk item has the costs $c$ then we decompose $c=\sum_i 2^i c_i$ and add an adder for $2^i$ conditioned on $c_i=1$. Therefore, a circuit for the addition of the costs on $n_c$ qubits consists of at most ${\rm log}(n_c)$ adders with $O(n_c)$ XOR gates each, which are controlled by at most $n_c$ qubits. In total, there are $O(n_r n_c{\rm log}n_c)$ XOR gates that are controlled by at most $n_c$ qubits.

To check whether the cost limit is reached or not we can simply calculate the difference between the accumulated costs and the limit and then just check the sign of the result. For this we calculate the binary two complement $c$ of the pre-defined limit and we write it in binary representation as $c=\sum_i 2^i c_i$. If $c_i=1$ then we append an adder for $2^i$ to the circuit. Therefore, we need at most $O({\rm log}(n_c))$ adders, which have $O(n_c)$ controlled XOR gates each.

To summarize, we see that the circuit for $n_r$ risk items, $n_t$ transitions (we assume that we converted many-to-one relations to the binary tree construction from above and that we have an additional overhead of ${\rm log}(n_r+n_t)$ extra qubits and operations) and $n_s \in O(n_r+n_t)$ control qubits consists of $O(n_r+n_t)$ gates that are controlled by at most $O({\rm log}(n_r+n_t))$ qubits each. Therefore, taking the overhead into account we need $O((n_r+n_t){\rm log}(n_r+n_t))$ gates for the circuit, which implements the model without costs. Since we need $O(n_r)$ adders with $O(n_c {\rm log}n_c)$ gates each we need $O(n_r n_c {\rm log}n_c)$ additional gates for the costs.

We now consider the complexity of the amplitude estimation for our model. The construction of the circuit can be seen in figure~\ref{modelinQAE}. We assume that the implementation of the risk model consists of $m_r \in O((n_r+n_t){\rm log}(n_r+n_t)+ (n_r n_c){\rm log}(n_c))$ gates that are at most controlled by $p$ gates each. The output of the estimation should have $n_{\rm ae}$ qubits. The circuit uses an $X$ gate and $n_{\rm ae}$ Hadamard gates as well as a Fourier transform on $n_{\rm ae}$ qubits, which needs $O(n_{\rm ae}^2)$ gates. The circuit uses the model once directly and also the simply controlled powers ${\rm QRM}^i$ for $i=1, \ldots, n_{\rm ae}$. Therefore, we need
$$
\sum_{i=1}^{n_{\rm ae}} 2^i m_r
$$
gates that are controlled by at most $p+1$ qubits each. Therefore, the circuit for the amplitude estimation is dominated by the controlled gates for the powers of the model and we need $O(2^{n_{\rm ae}}m_r)$ gates that are controlled by at most $p$ qubits each.

The circuit for the amplitude estimation is used inside the Grover search as shown in figure~\ref{grover}. This means, that for each iteration we need to implement the circuit for the amplitude estimation and for its inverse. If the amplitude estimation circuit needs $m_{\rm ae}$ gates that are controlled by at most $p$ qubits each then we need for $n_g$ Grover steps in total $2n_g\cdot m_{\rm ae}$ gates and the controlled phase gates and the gates on $O({\rm log}(n_r+n_t))$ qubits for the search for the modification we are looking for. If the register for the search space has $n_s$ qubits, then we need $O(n_{\rm ae}n_s)$ uncontrolled Hadamard and XOR gates and $n_{\rm ae}$ XOR gates that are controlled by $n_s-1$ qubits.\\
To illustrate what this means in practice, we consider a concrete example, as it might appear at Deutsche Börse Group. We take $n=150$ risk items, $n_t=250$ transition probabilities, $n_c=10$ qubits for the costs so the range of costs is from 0 to 1023 units, $n_{ae}=10$ qubit precision for the QAE and a search for one parameter which 
dominates the sensitivities. The calculations explained above then give
\begin{equation*}
2^{10}((150+250)\cdot{\rm log}(150+250) + 150\cdot 10 \cdot {\rm log}(10)) \approx 2.6 \cdot 10^{6}
\end{equation*}
gates for the QAE, and an additional factor of about $2\cdot 15$ from the Grover search (searching about 400 parameters requires about 15 Grover steps).\\
So the number of gates required for a typical real world application would be in the order of 100 million, many of which multiply controlled.\\
Several algorithms to decompose controlled gates into elementary gates are described in~\cite{barenco, cybenko}. If we do not use ancilla qubits, then a unitary on a qubit that is controlled by $n-1$ qubits can be implemented by $O(n^2)$ elementary gates on a register of $n$ qubits.

\section{Theory: Grover search with imperfect oracles}\label{theory}
In this section, we motivate our use of imperfect oracles in our Grover search as if they were perfect. We will do this by capturing the different kinds of deviations that our QAE based oracles exhibit in simple toy models, without using the actual QAE. We will then quantify the effects of the imperfections on the performance of the Grover search.

Some aspects of imperfect, unitary oracles were discussed in earlier work, see for example \cite{imperfectoracles}, \cite{biasedoracles1} and \cite{biasedoracles2}. For example, it was shown that an imperfect oracle can be used to build a new oracle which approximates a perfect one and one only incurs a constant factor in the complexity of the resulting circuit. Our aim is different: we want to show that simply using the imperfect oracle directly in Grover's algorithm leads to results that are good enough for our use case.\\
There are two types of deviations from perfect oracles: false positives, that is, the oracle marks a state that should not be marked, and false negatives, that is, the oracle does not mark a state that should be marked. The QAE oracles we built for our use case exhibit both, false positives and false negatives. The key question now is: how do false positives and false negatives affect the success probability of the Grover algorithm, and how do they affect the optimal number of search steps? In this section we will illustrate this effect with simple examples instead of examining the complicated QAE oracles. The result will be that having false negatives reduces the success probability of the search but not the number of optimal search steps, while false positives change the number of optimal search steps and the success probability.

\subsection{Oracles with false positives}\label{subsectionImperfectDetection}
A very simple way to construct unitary oracles which produce false positives is to take a perfect oracle, put a unitary transformation on the input qubits to mix good states with bad states, and put the inverse transformation after the application of the oracle.\\
To make things clear and concrete, we consider the case with a search space of size $N=2^n$ and one solution, $M=1$. We can always work in a basis where the solution is the basis vector represented by $|1\ldots 1\rangle$ on all $n$ qubits in the search space as in figure \ref{imperfectdetection}. There, the oracle acts on the search space (qubits $q_0$ to $q_{n-1})$ with an $(n-1)$-fold controlled NOT gate, which marks qubit $q_n$. The circuit produces a phase kickback on the search space, as required for the Grover algorithm. However, there are rotations on each of the input qubits, which rotate each input qubit by an angle $\alpha_i$. When all $\alpha_i$ are 0, the oracle perfectly recognizes the desired solution state $\ket{11\ldots1}$ and marks it. All other states are not marked.

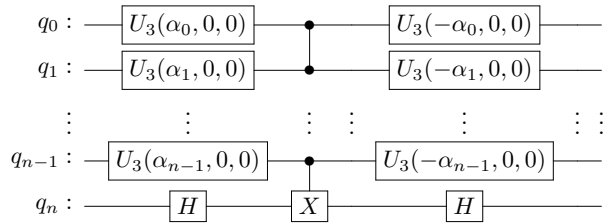
\begin{figure}[h]
\captionsetup{font=small,width=0.8\textwidth}
\centering
\scalebox{0.9}{
\Qcircuit @C=1.0em @R=0.3em @!R {
\lstick{ {q}_{0} :  } & \gate{U_3(\alpha_0,0,0)} & \ctrl{1} & \qw & \gate{U_3(-\alpha_0,0,0)} & \qw & \qw\\
\lstick{ {q}_{1} :  } & \gate{U_3(\alpha_1,0,0)} & \ctrl{0} & \qw & \gate{U_3(-\alpha_1,0,0)} & \qw & \qw \vspace{1cm} \\
\lstick{ \vdots } & \vdots                    & \vdots      & \vdots & \vdots                       & \vdots & \vdots \\
\lstick{ {q}_{n-1} :  } & \gate{U_3(\alpha_{n-1},0,0)} & \ctrl{1} & \qw & \gate{U_3(-\alpha_{n-1},0,0)} & \qw & \qw\\
\lstick{ {q}_{n} :  } & \gate{H}                 & \gate{X} & \qw & \gate{H}                  & \qw & \qw
}}
\caption{An example of an oracle acting on the search space $q_0$ to $q_{n-1}$. The internal qubit $q_n$ is marked and used to generate the phase flip on the search space. The imperfection of the oracle is parameterized by the angles $\alpha_i$. In the absence of imperfections (all $\alpha_i=0$) it flips the phase of qubit $q_n$ when the input qubits $q_0$ to $q_{n-1}$ are all in state $\ket{1}$.}\label{imperfectdetection}
\end{figure}

The effect of the rotations on the qubits is that the oracle will not only mark the true solution $\ket{1...1}$, but other states as well. The key observation here is that the imperfection of the oracle can be viewed as a change in the number of solutions from $M=1$ to $\hat{M}\neq1$. Consider, for example, a non-trivial $\alpha_0$. Then, when we count the solutions, summing over all input states, we find not only a contribution from $\ket{1...1}$, with weight ${\rm cos}(\alpha_0)$, but also one from $\ket{011...1}$, with weight $\sin(\alpha_0)$ and similarly for the other qubits. So for general angles $\alpha_i$ the number of solutions $\hat{M}$ is given by\footnote{Note that, depending on the value of $\alpha$, the number of effective solutions $\hat{M}$ can increase or decrease.}
\begin{equation} \label{solutions}
\hat{M} = \prod_{i=0}^{n-1} \left(\cos(\alpha_i) + \sin(\alpha_i)\right)^2
\end{equation}
For sufficiently large problems, we can use simple, classical results for Grover's algorithm (see, for example chapter 6 in \cite{nielsenchuang}). In particular, we find a solution with almost certainty if we set the number of search steps to\footnote{Note that according to equation \ref{solutions}, the imperfection may lower or increase the number of optimal steps in the search algorithm.}
\begin{eqnarray}
\hat{S} &=& \frac{\pi}{4}\sqrt{\frac{N}{\hat{M}}} \label{steps}
\end{eqnarray}
The success probability for Grover with perfect oracles is 1 (or almost 1) for sufficiently large search spaces. For our imperfect oracles, this is generally not the case. In our example, there is only one true solution $M=1$, but there are false solutions such that $\hat{M}>1$. The success probability $P$ is the probability to measure the true solution, rather than one of the fake solutions at the end of the circuit. The true solution is found with probability
\begin{equation}
P = \prod_{i=0}^{n-1} \cos(\alpha_i)^2  \label{success}
\end{equation}
To summarize, for the imperfect oracles in figure \ref{imperfectdetection}, we find a solution after $\hat{S}$ steps, given by equation \ref{steps} with the success probability $P$ in equation \ref{success}.\\
The transformations that we apply before and after the oracle are given by U3 gates on each input qubit in figure \ref{imperfectdetection}. We could also apply general unitary operators $U$ acting on a state vector with $N$ components. The relevant part of those is the mixing between the true solution state and the $i$-th state. So we can write $U = \prod_{i=1}^{n-1} R_i$, where $R_i$ is a rotation between the $N$-th and the $i$-th state. The number of solutions $\hat{M}$ is then given by summing up the projections of the resulting state on all states: $\sum_i \bra{i} U \ket{N}$.\\
The considerations above give some guidance on how the imperfections of our QAE oracles of section~\ref{quantumimplementation} take effect. For example, we could simulate the (at least) 81\% percent success probability of the QAE oracles by setting $\alpha_1/2=0.45$ (the factor of 1/2 again stemming from Qiskit conventions). According to formula~\ref{solutions} we get $1.8$ effective solutions which imply a reduction of Grover steps by about 25\% according to~\ref{steps}. The maximal success probability is 81\%, according to equation~\ref{success}. Note that the example we describe here does not only exhibit false positives, but also false negatives.\\
It is easy to see that the success probability that remains is still at least 81\% even if we distribute the imperfection over more qubits. However, it is also easy to see from equation~\ref{steps} that the optimal number of search steps can be affected significantly. One can engineer deviations by spreading the imperfection over many of the other qubits in our example. We want to stress that this is not a problem for our use case, however, since it corresponds to situations in which there does not exist a single parameter which has a much bigger impact on the problem than all others. Not finding such a parameter (that is, several runs of the algorithm yield different answers) then tells us that there is no dominantly sensitive parameter. In other words, if the overlap of the histograms produced by the QAE for each parameter is very small, then can assume that the success probability and the number of optimal search steps for our QAE oracles is that of perfect oracles.

\subsection{Oracles with false negatives}\label{subsectionImperfectMarking}
Now we equip the oracle with an internal state space, just like the QAE oracles have their internal state space onto which the QAE algorithm maps from the search space.\\
The internal state of the oracle will consist of $1 + k$ qubits. The first qubit will be switched on if a solution has been found. Each of the other $k$ qubits will be in a superposition of $\ket{0}$ and $\ket{1}$.
For the case $k=1$ the oracle can then be represented by the following circuit diagram:
\begin{equation}\label{circuitGroverMod}
\Qcircuit @C=1.0em @R=0.0em @!R{
\lstick{ {q}_{0} :  } & \qw       & \ctrl{1}             & \qw        & \ctrl{1}                 & \qw       & \qw     & \qw\\ 
\lstick{ {q}_{1} :  } & \qw       & \ctrl{2}             & \qw        & \ctrl{2}                 & \qw       & \qw     & \qw\\
\lstick{ {q}_{2} :  } & \qw  & \qw                  & \targ      & \qw                      & \qw  & \qw     & \qw \\
\lstick{ {q}_{3} :  } & \qw       & \multigate{1}{U} & \ctrl{1}   & \multigate{1}{U^{\dagger}} & \qw       & \qw     & \qw\\
\lstick{ {q}_{4} :  } & \qw       & \ghost{U}        & \ctrlo{-2} & \ghost{U^{\dagger}}       & \qw       & \qw     & \qw
}
\end{equation}
Here, $q_0$ and $q_1$ are the search space and the solution we search is the state $\ket{11}$. In this case, a perfect oracle should mark qubit $q_2$. The last two qubits $q_3$ and $q_4$ are the internal space of the oracle with $q_3$ indicating a solution. For the operator $U$ which acts on the last two qubits, we make the following choice: $U$ acts with a NOT gate on $q_3$ and with a Hadamard gate on $q_4$. This puts $q_4$ into the superposition $\sqrt{1/2}(\ket{0}+\ket{1})$. There is a second choice to build such an oracle, which is the same as that in \ref{circuitGroverMod} except that the doubly controlled NOT on $q_2$ is now activated by state $\ket{11}$ on qubits $q_3$ and $q_4$. Note that the circuit in \ref{circuitGroverMod} marks only partially, while the product of the two possible oracles, which we can write as 
\begin{equation*}
\Qcircuit @C=1.0em @R=0.0em @!R{
\lstick{ {q}_{0} :  } & \qw       & \ctrl{1}             & \qw           & \qw           & \ctrl{1}                 & \qw       & \qw     & \qw\\ 
\lstick{ {q}_{1} :  } & \qw       & \ctrl{2}             & \qw           & \qw           & \ctrl{2}                 & \qw       & \qw     & \qw\\
\lstick{ {q}_{2} :  } & \qw  & \qw                  & \targ         & \targ         & \qw                      & \qw  & \qw     & \qw \\
\lstick{ {q}_{3} :  } & \qw       & \multigate{1}{U} & \ctrl{1}      & \ctrl{1}     & \multigate{1}{U^{\dagger}} & \qw       & \qw     & \qw\\
\lstick{ {q}_{4} :  } & \qw       & \ghost{U}        & \ctrlo{-2}    & \ctrl{-2}     & \ghost{U_-^{\dagger}}       & \qw       & \qw     & \qw
}
\end{equation*}
marks perfectly. Of course, the internal space can be larger than one qubit. Then there are $2^k$ different oracles $\tilde{O}_i$, one for each of the possible controls. And again, it is easy to see that the product of all of them is simply a complicated implementation of the perfect oracle $O$, i.e., we have
\begin{equation}
O = \prod_{i=1}^{2^k} \tilde{O}_i
\end{equation}

Recall that in the oracles we built for our use case in section~\ref{sectionGroverSearch}, the output qubits of the QAE are used to control the phase operations in order to mark the states we are looking for. Usually, we will use several controlled NOT operations to mark states, because the relevant results of the QAE spread over two, four, six or even more combinations, which then together mark the state with at least 81\% probability. Therefore, our construction does not mark states perfectly, because we only use the most relevant outputs of QAE to control the phase, and we certainly might not use all relevant combinations for the marking. Products of the oracles described in this section work exactly the same way: here we also use controlled NOT with $2^k$ controls on the internal space of the oracle for the marking and hence we can use these oracles to study the effect of not fully marking solutions, just as it happens with the QAE oracles. So how does this imperfect marking affect the performance of the Grover search?\\
To investigate this, we first recall the standard Grover algorithm and introduce some notation. We work in the plane spanned by the equal superposition $\ket{s}$ of all $N$ states in the search space and the $M$ solutions $\ket{\omega}$:

\begin{equation}
\ket{s} = \frac{1}{\sqrt{N}} \sum_{i=0}^{N-1} \ket{x_i}
\end{equation}
such that
\begin{equation}
\braket{s | \omega} = \sqrt{\frac{M}{N}}
\end{equation}

We then define the operators
\begin{eqnarray}
U_\omega &=& I - 2 \ket{\omega}\bra{\omega}\\
U_s &=& (2 \ket{s}\bra{s} - I)
\end{eqnarray}

\noindent where, as is well known~\cite{aharonov}, $U_\omega$ is a reflection about $0$ (i.e., $\ket{\omega} \mapsto -\ket{\omega}$) and $U_s$ is a reflection about $s$. The product $G=U_sU_\omega$ of these two reflections then is a rotation in the plane spanned by $|s\rangle$ and $|\omega\rangle$ by a constant angle $\theta$, and $G^2$ is a rotation by $2\theta$ and so on. $G$ corresponds to one step of the Grover algorithm.\\
Now we want to replace the standard Grover operator with a perfect oracle by one with an imperfect oracle as described earlier in this section. We introduce the notation $|\pm\rangle=|0\rangle \pm |1\rangle$ in order to simplify the following calculations. When we enlarge the state space with $k$ ancilla qubits (since we cannot implement general roots of reflections otherwise), we can define the operators
\begin{eqnarray}
\tilde{U}_{\omega} &=& I - \frac{1}{2^{k-1}} \ket{\omega + \ldots +} \bra{\omega + \ldots +}\\
U_s &=& 2 \ket{s}\bra{s} \otimes I - I
\end{eqnarray}
There are another $2^{k-1}$ different $\tilde{U}_\omega$ operators, one for each combination of the $k$ states $|+\rangle$ and $|-\rangle$. However, we can perform the analysis without loss of generality with $+\ldots+$ and look at products of it afterwards to get to the general case.
We perform one step of the Grover algorithm with the Grover root operator:
\begin{equation}
\tilde{G} \ket{s0\ldots 0} = U_s \tilde{U}_\omega\ket{s0\ldots 0} = \ket{s 0\ldots 0} + \frac{1}{2^{k-1}\sqrt{N}} \ket{\omega+\ldots +} - \frac{2}{2^{k-1}N} \ket{s+\ldots +}
\end{equation}

Now assume we perform the Grover algorithm with a $\frac{a}{2^k}$-root Oracle, which is the product of $a$ different oracles, each with a different combination of plusses and minusses on the internal space. Then the increase in success probability $\Delta \tilde{P}_n$ for the Grover algorithm is given in terms of the regular Grover increase in success probability $\Delta P_n$ after $n$ steps by:

\begin{equation} \label{result2}
\Delta \tilde{P_n} = \frac{a}{2^k} \Delta P_n
\end{equation}
The proof is given in appendix \ref{proof}.\\
So the root Grover performs a rotation in the $\ket{s}$-$\ket{\omega}$-plane by the same angle $\theta$ as the standard Grover algorithm. The important point to note is that this means that the optimal number of Grover steps remains unchanged when we work with this kind of imperfect oracle. In applications, this means that if we do not tap all the relevant output states of the QAE in our QAE-oracles, success probability decreases, but the number of steps required remains constant.\\

\section{Conclusion}
We have demonstrated that the sensitivity analysis of a business risk model at Deutsche Börse Group can be implemented as a quantum program. Our results indicate that the expected theoretical quadratic speed advantage over the classical approach, consisting of running Monte Carlo simulations for each relevant set of parameters and finding the most sensitive set, can be attained.\\
The application of this model is not limited to business risk. For instance, credit risk and operational risk seem to be natural areas that could profit from the same quantum approach as well. More generally, the analysis of any model which consists of many random variables with relations resembling a tree structure could be a suitable problem for our approach. Since in general solutions based on classical Monte Carlo simulations seem the only viable solution for such problems~\footnote{Quasi-Monte Carlo simulations~\cite{fries2}, which in practice often scale approximately quadratically better than Monte Carlo simulations (in $s$ dimensions, it scales like $O(\frac{log(N)^s}{N})$), are not well suited to our problem, since it is high-dimensional, which destroys the advantage.}, this may become of practical importance as soon as the necessary quantum hardware is available.
Since the number of qubits that our model requires -- 200 would already suffice for our usecase -- is low compared to other quantum computing applications (as an example, the implementation of the Shor factoring algorithm for numbers of relevant size would require machines with more than an order of magnitude bigger as the key size should be at least 2048 bit for RSA), problems like this might be within the reach of the first generation of error corrected quantum computers.\\
At the heart of our contribution are our QAE-based oracles. They do not mark perfectly. Nevertheless, we use them directly in the Grover search and argued that we can still control the errors.\\
The theoretical part of the paper analyzed two types of imperfections in unitary oracles in the Grover search. The first type of imperfection is a mixing between states that are true solutions of the search problem and those that are not. We calculated the impact of the imperfection on the success probability of the search and on the optimal number of steps of the search. The second kind of imperfection corresponds to pure false negative detections of the oracle. For this kind of imperfection, we showed that the optimal number of search steps in the Grover algorithm remains unchanged. We also calculated the decrease in success probability of the search resulting from the imperfection.
\section{Acknowledgments}
We would like to thank Christoph Böhm, Michael Girg, Konrad Sippel, Christian Tüffers and Rory McLaren of Deutsche Börse Group for supporting this project.\\
The paper has benefited from comments by Fred Jendrzejewski and Philipp Hauke who kindly read a draft.\\
We would also like to thank Christian Fries, who explained the fact that in many cases, classical quasi-Monte Carlo simulations perform better than Monte Carlo simulations.\\
We acknowledge use of IBM Quantum for this work. \\
We thank NTT Global Data Centers EMEA, a division of NTT Ltd. for providing servers in the Technology Experience Lab in Frankfurt to conduct the simulations.
\appendix

\section{Proof details} \label{proof}
In this section, we provide the proof for equation~\ref{result2} in section~\ref{subsectionImperfectMarking}.
The key insight we use is that there is a space on which the usual Grover operator and the $2^k$th-root Grover operator act in exactly the same way. Each state from the case with the normal Grover oracle is simply replaced by the same state, but tensored with $\ket{+...+}$. We write $|+\rangle$ instead and work with the all-plus state without loss of generality.\\
The difference in the results between the usual and the root Grover algorithms stems from the first step in the algorithm. We calculate each step for the standard case and the imperfect case, where operators for the imperfect case are marked with a tilde. The Grover operators act on the initial state $\ket{s}$ like this

\begin{eqnarray}
G &=& U_s U_{\omega} \\
\tilde{G} &=& U_s \tilde{U}_{\omega}
\end{eqnarray}

We can then write
\begin{eqnarray}
G \ket{s} &=& \ket{s} + \ket{T}\\
\tilde{G} \ket{s 0} &=& \ket{s0} + \frac{1}{2^k} \ket{T +}
\end{eqnarray}
(where the form of $T$ is not important) and hence we have with
the definition 
$$Z=|T\rangle+G|T\rangle \ldots G^{n-1}|T\rangle$$ 
the equations
\begin{eqnarray}
G^n \ket{s} &=& \ket{s} + \ket{T} + G \ket{T} + ... + G^{n-1} \ket{T} \\
&=&  \ket{s} + \ket{Z}  \nonumber\\
\tilde{G}^n \ket{s 0} &=& \ket{s0}  + \frac{1}{2^k}\left( \ket{T +} + G\ket{T +} + ... + G^{n-1} \ket{T +} \right)\\
&=&  \ket{s0} + \frac{1}{2^k} \ket{Z+} \nonumber
\end{eqnarray}

The success probability $P_n$ for the usual Grover algorithm after $n$ steps is
\begin{eqnarray}
P_n &=& \bra{\omega} G^n \ket{s}^2 \\
&=& \left( \braket{\omega | s} + \braket{\omega|z} \right)^2 \nonumber\\
&=& \braket{\omega |s}^2 + 2 \braket{\omega | s} \braket{\omega | z} +  \braket{\omega | z}^2 \nonumber
\end{eqnarray}
The probability increase achieved by the Grover algorithm is
\begin{equation}
\Delta P_n = P_n - \braket{\omega |s}^2 = 2 \braket{\omega | s} \braket{\omega | z} +  \braket{\omega | z}^2
\end{equation}
We can now do the same thing for the root Grover, with the difference that we now sum over all ancilla states in the computational basis (denoted by $x$ with values from 0 to $2^k-1$):
\begin{eqnarray}
\tilde{P}_n &=& \sum_x \bra{\omega x} \tilde{G}^n \ket{s0}^2 \\
&=& \sum_x \left(\braket{\omega x|s0} + \frac{1}{2^k} \braket{\omega x | z+}\right)^2 \nonumber \\
&=& \left(\braket{\omega 0|s0} + \frac{1}{2^k} \braket{\omega 0 | z+}\right)^2 \nonumber\\
&& + \frac{1}{2^{2k}} \braket{\omega 1 | z+}^2 \nonumber \\
&&+ ... \nonumber \\
&&+ \frac{1}{2^{2k}} \braket{\omega (2^k-1) | z+}^2 \nonumber \\
&=& \braket{\omega 0|s0}^2 + \frac{2}{2^k} \braket{\omega x|s0} \braket{\omega 0|z+} \nonumber \\
&&+ \left( \frac{1}{2^k}\right)^2\sum_x \braket{\omega x| z+}^2 \nonumber \\
&=& \braket{\omega 0|s0}^2 + \frac{2}{2^k} \braket{\omega x |s0} \braket{\omega 0 |z+} \nonumber \\
&&+ \frac{1}{2^k} \braket{\omega x| z+}^2 \nonumber
\end{eqnarray}
The probability increase achieved by the root Grover algorithm is
\begin{eqnarray}
\Delta \tilde{P}_n = \tilde{P}_n - \braket{\omega 0 |s0}^2 &=& \frac{2}{2^k} \braket{\omega x |s0} \braket{\omega 0 |z+} + \frac{1}{2^k} \braket{\omega x| z+}^2
\end{eqnarray}
and hence we find that the relation between the probability increase $\Delta\tilde{P_n}$ of root Grover and the probability increase $\Delta P_n$ of Grover is
\begin{equation}
\Delta \tilde{P}_n = \tilde{P}_n - \braket{\omega_0 |s0}^2 = \frac{1}{2^k} \left( P_n - \braket{\omega |s}^2\right) = \frac{1}{2^k} \Delta P_n
\end{equation}

As a final step of the proof, we consider products of $a$ different oracles, each chosen from the $2^k$ different partial oracles. It is easy to see that the $2^k$ states in the basis consisting of $+$ and $-$ are orthogonal. The calculation above can be carried through in exactly the same way with a product of oracles. All products between different combinations of $+$ and $-$ cancel, and we simply end up summing up the result from a single oracle $a$ times. This concludes the proof.

\subsection*{Unequal activation}
In the last section, we went over the case where all the internal oracle states have an equal amplitude. Of course, this is generally not what we find with QAE. The $k$ qubit output can be thought of as a histogram of the $2^k$ states the qubits can describe. Luckily, the formalism described above goes through for the more general case with just minor amendments.\\
First, consider the case where the ancilla states are simply $\ket{+}$ and $\ket{-}$ as before. In general these can be transformed with a unitary. We will simplify this slightly and look at a rotation on a single qubit about an angle $\alpha$ instead. Then, instead of
\begin{eqnarray}
\braket{0|+} &=& 1\\
\braket{0|-} &=& 1\\
\braket{+|-} &=& 0
\end{eqnarray}
we have
\begin{eqnarray}
\braket{0| \hat{+}} &=& {\rm cos}(\alpha)\\
\braket{0| \hat{-}} &=& {\rm sin}(\alpha)\\
\braket{\hat{+}|\hat{-}} &=& 0
\end{eqnarray}
Note that this means that there is a difference in the success probabilities for the $+$ and the $-$ ancilla states. The factors carry through most of the calculation of the success probability. However, the final result does not factor nicely and we are left with
\begin{eqnarray}
P - \braket{\omega s|0}^2 &=& \frac{{\rm cos}(\alpha)}{2^k} \left( 2 \braket{\omega_i |s0} \braket{\omega_0 |z+} + {\rm cos}(\alpha)\braket{\omega_i |z+}^2  \right)\\
&>& \frac{{\rm cos}^2(\alpha)}{2^k} \left( 2 \braket{\omega_i |s0} \braket{\omega_0 |z+} + \braket{\omega_i |z+}^2  \right)
\end{eqnarray}
and so we have that the improvement in the probability $\tilde{P}$ is given by
\begin{equation}
\tilde{P} \geq \frac{{\rm cos}^2(\alpha)}{2^k} P
\end{equation}
The number of steps that the Grover algorithms needs is unchanged.\\
For bigger ancilla spaces, which are products of several $+$ and $-$, this works similarly, i.e., states with different combinations of $+$ and $-$ are still orthogonal. The projections of each state onto $\ket{0}$ will in general all be different, of course.


\begin{thebibliography}{99}

\bibitem{nielsenchuang}
\textit{Quantum Computation and Quantum Information}\\
Nielsen, M. and Chuang, I.L.\\
Cambridge University Press, 2010.

\bibitem{QAE} 
\textit{Quantum Amplitude Amplification and Estimation}\\
Brassard, G., Hoyer, P., Mosca, M., Tapp, A.\\
arXiv:quant-ph/0005055

\bibitem{shor}
\textit{Algorithms for quantum computation: discrete logarithms and factoring}\\
Shor, P.W. \\
Proceedings 35th Annual Symposium on Foundations of Computer Science. IEEE Comput. Soc. Press: 124–134.

\bibitem{grover}
\textit{A fast quantum mechanical algorithm for database search}\\
Grover, L.K.\\ 
Proceedings, 28th Annual ACM Symposium on the Theory of Computing, (May 1996) p. 212

\bibitem{grover2}
\textit{Quantum mechanics helps in searching for a needle in a haystack}\\
Grover, L.K.\\
Phys. Rev. Lett., 79(2):325,1997\\
arXiv:quant-ph/9706033

\bibitem{bbht98}
\textit{Tight bounds on quantum searching}\\
Boyer, M., Brassard, G., Hoyer, Pr. and Tapp, A.\\
Fortsch. Phys. -Prog.Phys., 46(4-5):493-505,1998

\bibitem{quantumriskanalysis}
\textit{Quantum Risk Analysis}\\
Woerner, S. and Egger, D.\\
arXiv:1806.06893 [quant-ph]

\bibitem{fries1}
Fries, C., private communication

\bibitem{fries2}
\textit{Mathematical Finance: Theory, Modeling, Implementation}\\
Fries, C.\\
Second Edition\\
to be published

\bibitem{quantumpricing0}
\textit{Quantum computational finance: Monte Carlo pricing of financial derivatives}\\
Rebentrost, P., Gupt, B. and Bromley, T.R.\\
arXiv:1805.00109

\bibitem{quantumoptionpricing}
\textit{Option Pricing using Quantum Computers}\\
Stamatopoulos,N., Egger, D.J., Sun,Y., Zoufal, C., Iten, R., Shen, N. and Woerner,S.\\
arXiv:1905.02666

\bibitem{aharonov}
\textit{Quantum Computation}
Aharonov, D.\\
In Stauffer, D. (editor)\\
Annual Reviews of Computational Physics VI\\
World Scientific, Singapore, 1999

\bibitem{faultygrover}
\textit{Impossibility of a Quantum Speed-up with a Faulty Oracle}\\
Regev, O. and Schiff, L.\\
arXiv:1202.1027v1[quant-ph]

\bibitem{walksearch}
\textit{Search via Quantum Walk}\\
Magniez, F. E., Nayak, A., Roland, J and Santha, M.\\
arXiv:quant-ph/0608026

\bibitem{akr}
\textit{Coins Make Quantum Walks Faster}\\
Ambainis, A., Kempe, J. and Rivosh, A.\\
arXiv:quant-ph/0402107

\bibitem{ambianis}
\textit{Quantum Search Algorithms}\\
Ambainis, A.\\
arXiv:quant-ph/0504012

\bibitem{tulsi}
\textit{General framework for quantum search algorithms}\\
Tulsi, A.\\
arXiv:quant-ph/0806.1257

\bibitem{qiskit}
\textit{Qiskit: An Open-source Framework for Quantum Computing}\\
Abraham, H. et al.\\
2019\\
doi: 10.5281/zenodo.2562110


\bibitem{imperfectoracles}
\textit{Quantum Search on Bounded Error Inputs}\\
Høyer, P., Mosca, M. and de Wolf, R\\
arXiv:quant-ph/0304052v2

\bibitem{biasedoracles1}
\textit{General Bounds of Quantum Biased Oracles}\\
Iwama, K., Raymond, R. and Yamashita, S.\\
IPSJ Digital Courier, 1:415-425, 2005

\bibitem{biasedoracles2}
\textit{Robust Quantum Algorithms with $\epsilon$-Biased Oracles}\\
Suzuki, T., Yamashita, S., Masaki Nakanishi Katsumasa\\
arXiv:quant-ph/0605077

\bibitem{barenco}
\textit{Elementary gates for quantum computation}\\
A. Barenco, C.H. Bennett, R. Cleve, D.P. DiVincenzo, N. Margolus, P. Shor, T. Sleator, J. Smolin, H. Weinfurter\\
Physical Review A, 52:3457-3467, 1995.\\
arXiv:quant-ph/9503016

\bibitem{cybenko}
\textit{Reducing quantum computations to elementary unitary operations}\\
G. Cybenko\\
Computing in Science and Engineering 3(2):27 - 32, 2001.
\end{thebibliography}
\end{document}